\documentclass[letter,12pt]{article}%
\usepackage{amsmath,bm}
\usepackage{amsfonts}
\usepackage{amssymb}
\usepackage{graphicx}
\usepackage{rotating}
\usepackage[height=9in,left=1in,right=0.75in,bottom=1in]{geometry}
\usepackage[breaklinks=true,bookmarksopen=true,colorlinks=true,citecolor=blue]%
{hyperref}
\usepackage[authoryear]{natbib}
\usepackage{color}
\usepackage{colortbl}
\usepackage{paralist}
\usepackage{amsmath}%
\providecommand{\U}[1]{\protect\rule{.1in}{.1in}}
\RequirePackage{hypernat}
\newtheorem{theorem}{Theorem}[section]

\newtheorem{assumption}{Assumption}

\newtheorem{corollary}{Corollary}[section]

\newtheorem{definition}{Definition}[section]

\newtheorem{lemma}[theorem]{Lemma}

\newtheorem{proposition}{Proposition}[section]
\newtheorem{remark}{Remark}[section]

\catcode`~=11
\newcommand{\urltilde}{\kern -.15em\lower .7ex\hbox{~}\kern .04em}
\catcode`~=13
\def \@seccntformat#1{\csname the#1\endcsname.\quad}
\makeatother
\hypersetup{pdftitle={Escanciano}, pdfsubject={Semiparametric Identification and Fisher Information}, pdfauthor={Juan Carlos Escanciano}, pdfkeywords={Identification; Irregular Identification; Semiparametric Models; Fisher Information} }
\begin{document}

\title{Irregular Identification of Structural Models with Nonparametric Unobserved Heterogeneity}
\author{Juan Carlos Escanciano\thanks{Department of Economics, Universidad Carlos III
de Madrid, Calle Madrid 126, Getafe 28907, Madrid, Spain. E-mail:
\href{mailto:jescanci@indiana.edu}{jescanci@eco.uc3m.es}. Web Page:
\href{https://sites.google.com/view/juancarlosescanciano}{https://sites.google.com/view/juancarlosescanciano}%
. Research funded by the Spanish Grant PGC2018-096732-B-I00.}\\\textit{Universidad Carlos III de Madrid}}
\date{May 18th, 2020}
\maketitle

\begin{abstract}
One of the most important empirical findings in microeconometrics is the
pervasiveness of heterogeneity in economic behaviour (cf. Heckman 2001). This
paper shows that cumulative distribution functions and quantiles of the
nonparametric unobserved heterogeneity have an infinite efficiency bound in
many structural economic models of interest. The paper presents a relatively
simple check of this fact. The usefulness of the theory is demonstrated with
several relevant examples in economics, including, among others, the
proportion of individuals with severe long term unemployment duration, the
average marginal effect and the proportion of individuals with a positive
marginal effect in a correlated random coefficient model with heterogenous
first-stage effects, and the distribution and quantiles of random coefficients
in linear, binary and the Mixed Logit models. Monte Carlo simulations
illustrate the finite sample implications of our findings for the distribution
and quantiles of the random coefficients in the Mixed Logit model.

\vspace{2mm}

\begin{description}
\item[Keywords:] Irregular Identification; Semiparametric Models;
Nonparametric Unobserved Heterogeneity.

\item[\emph{JEL classification:}] C14; C31; C33; C35\newpage

\end{description}
\end{abstract}

\section{Introduction}

A tenet in empirical microeconometrics research is the pervasiveness of
heterogeneity in behaviour of otherwise observationally equivalent individuals
(cf. Heckman 2001). This paper shows that, for a large class of structural
economic models, regular identification of functionals of nonparametric
unobserved heterogeneity (UH), that is, identification of these functionals
with a finite efficiency bound, implies certain \textit{necessary} smoothness
conditions on the functional, leading to a practically simple check for
regularity (or lack thereof). In particular, this paper uses these
implications to show that cumulative distribution functions (CDFs) and
quantiles of UH often have infinite efficiency bounds in many empirically
relevant economic models with nonparametric UH. These results have important
practical implications, as these parameters are relevant for policy analysis,
and they explain why any inferences on such parameters are expected to be
unstable in empirical work. In particular, if a parameter is irregularly
identified, then no regular estimator with a parametric rate of convergence
exists (see Chamberlain 1986).

These observations are applicable to a wide class of models with nonparametric
UH. We consider first continuous mixtures, which have been commonly employed
as a modeling device to account for UH in a variety of economic settings
ranging from labour to industrial organization; see Compiani and Kitamura
(2016) for a recent review. The canonical example is a tightly specified
structural \emph{parametric} model that is made flexible by allowing all (or a
subset) of parameters to be individual specific, thereby accounting for UH. We
show that if the mapping from the individual specific parameters to the
conditional likelihood is smooth, then there will be many functionals of UH
that will not be regularly identified. Heuristically, smoothness of the
conditional likelihood translates into a multicollinearity problem, as we
further explain below. There are important economic applications that fall
under this setting, see, e.g., Heckman and Singer (1984a, 1984b) for the study
of unemployment duration. We demonstrate the usefulness of these results in
the context of duration data by establishing an infinite efficiency bound for
the distribution and quantiles of UH in the structural model of unemployment
duration with two spells and nonparametric UH recently proposed by Alvarez,
Borovickov\'{a} and Shimer (2016).

The results are then extended to several classes of Random Coefficients (RC)
models. These models have a long history in economics; see, e.g., Masten
(2017) for a review of the literature. Applying our results to these models is
technically more involved because these models have discontinuous conditional
likelihoods given UH. We consider first RC models where UH is independent of
regressors and establish an infinite efficiency bound for the distribution and
quantiles of UH in binary and linear RC models. Establishing the zero
information in the linear RC model is particularly challenging because the
discontinuity in the conditional likelihood leads to potential discontinuities
in the scores of the model. Given these results, we extend them to a
triangular RC model with a continuous endogenous variable, where we show
irregular identification of the average marginal effect (AME) and the
proportion of individuals with a positive marginal effect. The irregularity of
the AME is driven by a positive mass of individuals with small first-stage
effects. The irregular identification of the CDF and quantiles of the
distribution of random or correlated effects holds more generally.

The models treated up to this point are indexed by the distribution of UH, and
only by that distribution. However, a simple and powerful observation of this
paper is that our analysis can be trivially extended to more complex
semiparametric models indexed by UH and additional (possibly
infinite-dimensional) parameters. We illustrate this point with several
examples, including semiparametric mixture models where some parameters are
fixed and others are random. A leading example is the popular RC Logit or
Mixed Logit model, which is one of the most commonly used models in applied
choice analysis. This model was introduced by Boyd and Mellman (1980) and
Cardell and Dunbar (1980) and it is widely used in environmental economics,
industrial economics, marketing, public economics, transportation economics
and other fields. Applying our results to this model we obtain an infinite
efficiency bound for CDFs and quantiles of the RC. The Mixed Logit example
nicely illustrates the most appealing feature of our method of proof, which is
its simplicity. Two lines of proof and a simple application of dominated
convergence suffice. This should be contrasted with direct efficiency bounds
calculations, which are particularly challenging for this model (or for any of
the models we consider for that matter). These results have practical
implications for proposed estimators of the Mixed Logit model. We report Monte
Carlo simulations supporting our theoretical findings for \textquotedblleft
fixed grid\textquotedblright\ estimators of the distribution and quantiles in
the Mixed Logit model (cf. Bajari, Fox and Ryan 2007 and Fox, Kim and Yang
2016). Further illustrations demonstrating the utility of our results in
semiparametric settings are gathered in an Appendix and include examples on
mixed proportional duration models and measurement error models with two
measurements identified by means of Kotlarski's lemma.

The parameters (functionals) we consider are of interest in their own. For
example, labour economists are interested in the proportion of individuals at
risk of severe long term unemployment, and more generally, social scientists
are interested in evaluating the effects of treatments and policy
interventions (e.g. average marginal effects and average signs). The
functionals that we entertain, such as CDFs and quantiles of UH, are also used
as imputs in subsequent counterfactual exercises. Our research limits the kind
of inferences that are attainable with these parameters in models where UH is nonparametric.

What can be done to obtain regular identification of CDFs and quantiles of UH
in these models? We show in several examples that functional form assumptions
that restrict the conditional likelihood of observables given heterogeneity do
not generally help for the purpose of achieving regularity of quantiles and
CDFs if UH is still nonparametric. Thus, our results show that restricting UH
is somewhat necessary to attain finite efficiency bounds for the distribution
and quantiles of UH in many of the aforementioned models. Commonly used
strategies in practice, such as the use of parametric distributions for UH or
considering discrete heterogeneity, indeed restore the regular identification
of functionals of UH but can be deemed too strong. We find necessary
conditions of regular identification under semiparametric restrictions on UH,
although we recognize that giving general primitive assumptions for these
conditions seems difficult. Our recommendation for inference on CDFs and
quantiles of UH is to use flexible semiparametric specifications such as sieve
methods; see, e.g., Shen (1997), Chen (2007), Bajari, Fox and Ryan (2007), Hu
and Schennach (2008), Bester and Hansen (2007), Chen and Liao (2014), Fox, Kim
and Yang (2016) and references therein, coupled with regularization
(penalization) to reduce the high variance of estimates of functionals of UH
when the conditional likelihood is a very smooth function of UH, as
illustrated in this paper with the Mixed Logit model.

The rest of the paper is organized as follows. After a literature review,
Section \ref{Setting} sets notation and considers the class of continuous
mixtures, where the method is most transparent. This section illustrates the
theoretical results in the structural model of Alvarez, Borovickov\'{a} and
Shimer (2016). Section \ref{RC} extends the analysis to several classes of RC
models. Section \ref{SemiparametricModels} extends further the analysis to
semiparametric models, illustrating the theory with the Mixed Logit model.
Section \ref{Reg} discusses different strategies, some of them considered in
the literature, to regularize the estimation of CDFs and quantiles of UH.
Section \ref{MC} reports the results of some Monte Carlo simulations for the
CDF and quantiles of the distribution of UH in the Mixed Logit model. Section
\ref{Conclusions} concludes. An Appendix contains proofs of the main results,
further results on nonlinear RC models, examples and simulations.

\section{Literature Review}

Our paper relates to a number of studies providing sufficient conditions for
nonparametric identification for the distribution of UH in the aforementioned
models. See, among many others, Elbers and Ridder (1982), Heckman and Singer
(1984a, 1984b) and Alvarez, Borovickov\'{a} and Shimer (2016) for structural
models of unemployment duration, Beran and Hall (1992), Beran, Feuerverger and
Hall (1996), and Hoderlein, Klemela and Mammen (2010) for linear RC, Ichimura
and Thompson (1998), Gautier and Kitamura (2013) and Hoderlein and Sherman
(2015) for binary RC, Briesch, Chintagunta and Matzkin (2010) and Fox, Kim,
Ryan and Bajari (2012) for RC multinomial choice models, Hoderlein, Holzmann
and Meister (2017) for triangular RC models, Masten (2017) for simultaneous RC
models, and Lewbel and Pendakur (2017) for nonlinear RC models. For a review
of nonparametric identification results see Matzkin (2007, 2013) and Lewbel
(2019). What differentiates our paper from these and other related studies is
our focus on establishing whether identification is regular or not.

Establishing an infinite efficiency bound for functionals of UH in these
models is a priori a rather challenging task. The main reason is that
characterizing the so-called tangent space of the model and projections onto
it is generally quite complicated in the models we study here, and it may
explain the relative lack of theoretical work on semiparametric efficiency
bounds in RC and related models. See Newey (1990) for a review of
semiparametric efficiency bounds and some of the related concepts. Our method
of proof avoids the complications in directly computing the tangent space,
projections and the Fisher information, which is the standard approach in the
literature for obtaining efficiency bounds (see, e.g., Chamberlain 1986, Khan
and Tamer 2010). Our indirect method of proof is relatively much simpler. The
basic tool is a dominated convergence theorem, with regularity conditions that
are easy to check in many models (although not in all models). The main
building block is a fundamental result by van der Vaart (1991), who found a
necessary condition for regular estimation of a parameter. The main
observation of our paper consists in systematically exploiting the
implications that van der Vaart's (1991) necessary condition has on the
smoothness of certain influence functions. van der Vaart (1991), Groeneboom
and Wellner (1992) and Bickel, Klassen, Ritov and Wellner (1998) have also
used the necessary condition of van der Vaart (1991) to show that CDFs are
irregularly identified in some specific univariate exponential and uniform
mixture models. Relative to this work, our contribution is to derive
sufficient conditions for a general method of proof, thereby extending the
scope of applications to models of economic interest. In particular, we allow
for multidimensional UH, semiparametric models and non-smooth conditional
likelihoods such as those that arise with RC models.

Although not the focus of this paper, a large class of models for which our
results are applicable are panel data models with fixed effects. Within this
setting, Chamberlain (1992) established regular identification of the AME in a
linear RC panel data model, while Arellano and Bonhomme (2012) showed the
identification of the full distribution of UH in a model with limited serial
dependence in errors. Graham and Powell (2012) pointed out the irregular
identification of the AME when regressors exhibit little variation across
periods, while Bonhomme (2011) derived conditions for regular and irregular
identification of moments of UH in nonlinear panel data. Our research is
highly complementary to these papers, as we consider different models and our
approach for proving irregular identification is different and exploits the
smoothness implications of regular identification.

We illustrate the theoretical results with some Monte Carlo simulations
implementing the \textquotedblleft fixed grid\textquotedblright\ nonparametric
CDF estimator of Bajari, Fox and Ryan (2007) and Fox, Kim, Ryan and Bajari
(2011), and further investigated in Fox, Kim and Yang (2016). We contribute to
the literature on the Mixed Logit model by proving the infinite efficiency
bound for the CDF and quantiles of the nonparametric distribution of RC. We
report further finite sample evidence on the performance of their
computationally attractive \textquotedblleft fix grid\textquotedblright%
\ estimator for CDFs and quantiles, as well as some regularized variants,
complementing recent work in econometrics by Horowitz and Nesheim (2019) and
Heiss, Hetzenecker and Osterhaus (2019).

\section{Basic Setting and Results}

\label{Setting}

Let $\{(Z_{i},\alpha_{i})\}_{i=1}^{n}$ denote an independent and identically
distributed (iid) sample with the same distribution as $(Z,\alpha)$. The
observed data is $Z_{1},...,Z_{n},$ while $\alpha_{i}$ denotes the $i$-th
individual's UH. Assume each observation $Z_{i}$ has a probability law
$\mathbb{P}$ and a density with respect to (wrt) a $\sigma-$finite measure
$\mu$ given by
\begin{equation}
f_{\eta_{0}}(z)=\int_{\mathcal{A}}f_{z/\alpha}(z)d\eta_{0}(\alpha),
\label{mixture}%
\end{equation}
where $f_{z/\alpha}(z)$ denotes the known conditional density of $Z$ given
$\alpha,$ and $\eta_{0}$ is the unknown distribution of $\alpha$ with support
on $\mathcal{A}\subseteq\mathbb{R}^{d_{\alpha}}$ (the results can potentially
be extended to abstract heterogeneity spaces, but for simplicity of exposition
we focus on the Euclidean case). The assumption of known conditional density
$f_{z/\alpha}(z)$ is relaxed in Section \ref{SemiparametricModels}.

Suppose we are interested in estimating a moment of UH,%
\[
\phi(\eta_{0})=\mathbb{E}_{\eta_{0}}[r(\alpha)],
\]
for a measurable function $r\left(  \cdot\right)  \in L_{2}(\eta_{0}),$ where,
henceforth, $\mathbb{E}_{\eta_{0}}$ denotes the expectation under the
distribution $\eta_{0}\ $and $L_{p}(\nu)$ denotes the space of (equivalence
classes of) real-valued measurable functions $h$ such that $\int\left\vert
h\right\vert ^{p}d\nu<\infty,$ for a generic measure $\nu.$ Henceforth, we
drop the sets of integration in integrals and the qualification $\nu-$almost
surely for simplicity of notation$.$ So, for example, a function in $L_{2}%
(\nu)$ is discontinuous when there is no continuous function in its
equivalence class. Also, we drop the reference to the measure $\nu$ in
$L_{2}(\nu)$ when $\nu=\mathbb{P}$, and write simply $L_{2}$. We will be
concerned with regular identification of $\phi(\eta_{0}),$ i.e. identification
of $\phi(\eta_{0})$ with a finite efficiency bound, when UH is nonparametric
as formally defined below.

The basic message of this paper is based on two observations. First, from a
general result in van der Vaart (1991), we prove that a necessary condition
for regular identification of $\phi(\eta_{0})$ when UH is nonparametric is the
existence of a measurable function $s(Z)$ with zero mean and finite variance
such that%
\begin{equation}
r(\alpha)-\phi(\eta_{0})=\int s(z)f_{z/\alpha}(z)d\mu(z).\label{reg}%
\end{equation}
Second, if the mapping $\alpha\rightarrow f_{z/\alpha}$ is continuous
(smooth), then under mild regularity conditions, (\ref{reg}) implies that
$r(\cdot)$ must be also continuous (smooth)$.$ The bulk of this paper is a
formalization of the second observation and its application to some economic
models of interest.

The precise sense of UH being nonparametric is the usual one, formalized as
follows. Let $H$ denote a class of distributions on $\mathcal{A}$, and assume
$\eta_{0}\in H$. Let $\eta_{t}\in H$ be a parametric submodel indexed by
$t\in\lbrack0,\varepsilon),$ for some $\varepsilon>0,$ such that for a $b\in
L_{2}(\eta_{0})$ the classical mean square differentiability condition holds,
\begin{equation}
\int\left[  \frac{d\eta_{t}^{1/2}-d\eta_{0}^{1/2}}{t}-\frac{1}{2}bd\eta
_{0}^{1/2}\right]  ^{2}\rightarrow0\text{ as }t\downarrow0.\label{0}%
\end{equation}
Then, a formal definition of nonparametric UH is given as follows. Denote by
$T(\eta_{0})$ the linear span of the $b^{\prime}s$ in (\ref{0}) and let
$L_{2}^{0}(\nu)$ denote the subspace of functions in $L_{2}(\nu)$ with zero
$\nu-$mean.

\begin{definition}
UH is nonparametric if $T(\eta_{0})$ is dense in $L_{2}^{0}(\eta_{0})$.
\end{definition}

Henceforth, we assume, unless otherwise stated, that UH is nonparametric. The
first result in this section, which follows from an application of van der
Vaart (1991), shows that, in the presence of nonparametric UH in model
(\ref{mixture}), regular identification of $\mathbb{E}_{\eta_{0}}[r(\alpha)]$
requires necessarily that (\ref{reg}) holds.

\begin{lemma}
\label{Regu}If UH is nonparametric, then (\ref{reg}) is necessary for regular
identification of $\phi(\eta_{0}).$
\end{lemma}

We note that Severini and Tripathi (2006, 2012) and Bonhomme (2011) have found
related results in the context of nonparametric instrumental variables and
nonlinear panel data models, respectively. Also, Escanciano (2020) has shown
that (\ref{reg}) is also sufficient for semiparametric identification of
$\phi(\eta_{0})$ in model (\ref{mixture}). Note that we are not assuming here
that $\eta_{0}$ or $s$ in (\ref{reg}) are identified. This generality is
important because these functions may not be identified in many structural
economics models under weak assumptions, which does not prevent us from
identifying and estimating certain functionals of them (cf. Hurwicz
1950).\footnote{Of course, if $\eta_{0}$ is identified, so is $\phi(\eta_{0})$
(since $r$ is known). Identification of $\phi(\eta_{0})$ follows from
(\ref{reg}) because we can find an identified function $\tilde{s}(Z),$
depending only on $f_{z/\alpha}$ and $r,$ such that $r(\alpha)=\mathbb{E}%
\left[  \left.  \tilde{s}(Z)\right\vert \alpha\right]  $ holds, and thus by
iterated expectations $\phi(\eta_{0})=\mathbb{E}_{\eta_{0}}[r(\alpha
)]=\mathbb{E}_{\eta_{0}}[\mathbb{E}\left[  \left.  \tilde{s}(Z)\right\vert
\alpha\right]  ]=\mathbb{E}\left[  \tilde{s}(Z)\right]  .$}

We now proceed with the main insight of this paper, which is that if the
mapping $\alpha\rightarrow f_{z/\alpha}$ is continuous (smooth), then, under
regularity conditions, $r(\cdot)$ must be also continuous (smooth)$.$ This
simple observation follows by dominated convergence, and it implies
non-regularity of CDFs, signs, quantiles, and other functionals of UH in
\textquotedblleft smooth models\textquotedblright\ satisfying the following
assumption. Let $N$ denote an open subset of $\mathcal{A}\subset
\mathbb{R}^{d_{\alpha}}.$

\begin{assumption}
\label{smooth} (i) $\alpha\rightarrow f_{z/\alpha}(z)$ is continuous on $N$
a.e-$\mu;$ (ii) for all $\alpha\in N$ there exists a neighborhood of $\alpha,$
say $\Gamma_{0}\subset N,$ such that for all $s$ satisfying (\ref{reg}),
\begin{equation}
\int\left\vert s(z)\right\vert \sup_{\alpha\in\Gamma_{0}}f_{z/\alpha}%
(z)d\mu(z)<\infty. \label{dominance}%
\end{equation}

\end{assumption}

Assumption \ref{smooth}(i) is easy to check. Assumption \ref{smooth}(ii) is a
dominance condition. The main complication in checking Assumption
\ref{smooth}(ii) is that $s$ belongs to $L_{2}(\mathbb{P})$ but not
necessarily to $L_{1}(\mu)$ or $L_{2}(\mu)$. We verify these conditions in a
number of examples below.

\begin{lemma}
\label{Main}Let the conditional density $f_{z/\alpha}(z)$ satisfy Assumption
\ref{smooth}. Then, $r(\alpha)$ in (\ref{reg}) is continuous in $\alpha$ on
$N.$
\end{lemma}

\noindent The following corollary is a direct consequence of the previous two lemmas.

\begin{corollary}
\label{Corcdf}Let Assumption \ref{smooth} hold. The CDF $\phi(\eta
_{0})=\mathbb{E}_{\eta_{0}}[1(\alpha\leq\alpha_{r})],$ for $\alpha_{r}\in N,$
is not regularly identified.
\end{corollary}

Quantiles of UH are nonlinear functionals, and are not covered by the previous
results. To extend the theory to a more general setting including nonlinear
functionals we need to introduce some notation. A functional $\phi(\eta
_{0}):H\rightarrow\mathbb{R}$ is said to be differentiable if there exists an
$r_{\phi}\in L_{2}^{0}(\eta_{0})$ such that for all paths satisfying
(\ref{0}), it holds%
\[
\lim_{t\rightarrow0}\frac{\phi(\eta_{t})-\phi(\eta_{0})}{t}=\mathbb{E}%
_{\eta_{0}}[r_{\phi}(\alpha)b(\alpha)].
\]
Under nonparametric UH such $r_{\phi}$ is unique, as in Newey (1994). This
function $r_{\phi}$ plays the role of the preceding moment function $r.$

To illustrate with an example, consider the scalar UH case and assume
$\eta_{0}$ is absolute continuous with a strictly positive Lebesgue density in
a neighborhood of $\phi(\eta_{0}),$ where $\phi(\eta_{0})$ is such that
\begin{equation}
\int_{-\infty}^{\phi(\eta_{0})}d\eta_{0}(\alpha)=\tau,\text{ }\tau\in(0,1).
\label{q1}%
\end{equation}
That is, $\phi(\eta_{0})$ is the $\tau$-quantile of $\eta_{0}$. It is
well-known, see, e.g., Lemma 21.3 in van der Vaart (1998), that the quantile
functional is differentiable under the conditions above with influence
function
\[
r_{\phi}(\alpha)=\frac{-\left\{  1(\alpha<\phi(\eta_{0}))-\tau\right\}  }%
{\dot{\eta}_{0}(\phi(\eta_{0}))},
\]
where $\dot{\eta}_{0}$ is the density pertaining to $\eta_{0}$. From our
results, the discontinuity of the influence function $r_{\phi}(\cdot)$ implies
irregular identification. Next result, formalizes this finding.

\begin{corollary}
\label{Corq}Let Assumption \ref{smooth} hold. Assume $\eta_{0}$ is absolute
continuous with a strictly positive Lebesgue density in a neighborhood of
$\phi(\eta_{0})$ satisfying (\ref{q1}). If $\phi(\eta_{0})\in N,$ then the
$\tau$-quantile of the nonparametric UH distribution is not regularly identified.

\begin{remark}
Henceforth, whenever we discuss identification of quantiles, we implicitly
assume that the components of UH have densities that satisfy the conditions in
Corollary \ref{Corq}. This example illustrates how our results are applicable
to nonlinear differentiable functionals.
\end{remark}
\end{corollary}

We discuss now the complications of the more standard approach of computing
the Fisher Information or the efficiency bound. Define the so-called tangent
space of scores $\mathcal{S}:=\{s\in L_{2}^{0}:s(z)=\mathbb{E}\left[  \left.
b(\alpha)\right\vert Z\right]  $ for some $b\in T(\eta_{0})\}.$ Then, a
standard result in linear inverse problems is that all solutions $s$ of
equation (\ref{reg}) have the same orthogonal projection onto the closure of
$\mathcal{S}$ (see Engl, Hanke and Nuebauer, 1996). Denote by $s^{\ast}$ such
orthogonal projection, the so-called efficient score. The efficiency bound is
given by the variance of $s^{\ast}(Z)$ (see e.g. Newey 1990, van der Vaart
1998, Bickel et al. 1998, and Escanciano 2020). Thus, an alternative to our
approach is to compute $s^{\ast}(Z)$ and checking that it has infinite
variance. However, computing $s^{\ast}(Z)$ can be cumbersome, particularly
because characterizing the mean squared closure of $\mathcal{S}$ can be a
rather difficult task in the models we analyze here. In fact, to the best of
our knowledge, the analytical expression for $s^{\ast}$ remains unknown for
the functionals and models we study. In passing, we note that these arguments
show that it suffices to check the dominance condition (\ref{dominance}) for
$s$ in the closure of $\mathcal{S}$. This additional information will turn out
to be quite useful in some of our applications, such as the linear RC model.

\subsection{An Application To A Structural Model of Unemployment}

We illustrate the applicability of the previous results in the context of a
structural model of unemployment with nonparametric UH. Nonparametric
heterogeneity has played a critical role in rationalizing unemployment
duration ever since the seminal contributions by Elbers and Ridder (1982) and
Heckman and Singer (1984a, 1984b). Recent work by Alvarez et al. (2016) is
motivated from this perspective. These authors have shown nonparametric
identification of the distribution of UH in their nonparametric structural
model for unemployment with two spells. Specifically, Alvarez, Borovickov\'{a}
and Shimer (2016) propose a structural model for transitions in and out of
employment that implies a duration of unemployment given by the first passage
time of a Brownian motion with drift, a random variable with an inverse
Gaussian distribution. The parameters of the inverse Gaussian distribution are
allowed to vary in arbitrary ways to account for UH in workers. These authors
investigate nonparametric identification of the distribution of UH, $\eta
_{0},$ when two unemployment spells $Z_{i}=(t_{i1},t_{i2})$ are observed on
the set $\mathcal{T}^{2},$ $\mathcal{T}\subseteq\lbrack0,\infty)$. The reduced
form parameters $\alpha=(\alpha_{1},\alpha_{2})^{\prime}\in\mathbb{R}%
\times\lbrack0,\infty)$ are functions of structural parameters. The
distribution of $Z_{i}$ is absolutely continuous with Lebesgue density
$f_{\eta_{0}}(t_{1},t_{2})$ given, up to a normalizing constant, by%
\begin{equation}
f_{\eta_{0}}(t_{1},t_{2})=\int_{\mathbb{R}\times\lbrack0,\infty)}\frac
{\alpha_{2}^{2}}{t_{1}^{3/2}t_{2}^{3/2}}e^{-\frac{\left(  \alpha_{1}%
t_{1}-\alpha_{2}\right)  ^{2}}{2t_{1}}-\frac{\left(  \alpha_{1}t_{2}%
-\alpha_{2}\right)  ^{2}}{2t_{2}}}d\eta_{0}(\alpha_{1},\alpha_{2}%
).\label{densityEx1}%
\end{equation}
Alvarez, Borovickov\'{a} and Shimer (2016) show that $\eta_{0}$ is
nonparametrically identified up to the sign of $\alpha_{1},$ but they do not
investigate if specific functionals of this distribution are regularly or
irregularly identified, which is the focus of study here. Specifically, we
show that the CDF of $\eta_{0}$ at a point, and other functionals of $\eta
_{0}$ with discontinuous influence functions, such as quantiles, have infinite
efficiency bounds. These functionals are important parameters. For example,
$\phi(\eta_{0})=\mathbb{E}_{\eta_{0}}\left[  1\left(  \alpha_{1}\leq
\alpha_{10}\right)  1\left(  \alpha_{2}\leq\alpha_{20}\right)  \right]  ,$ for
a fixed $\alpha_{10}<0<\alpha_{20}$ and large absolute values of $\alpha_{10}$
and $\alpha_{20},$ quantifies the proportion of individuals at risk of severe
long term unemployment (an individual with parameters $\alpha_{1}$ and
$\alpha_{2},$ $\alpha_{1}\leq\alpha_{10}$ and $\alpha_{2}\leq\alpha_{20},$ has
a probability larger or equal than $1-\exp(2\alpha_{10}\alpha_{20})$ of
remaining unemployed forever). We apply our previous results to this example
for a generic moment $\phi(\eta_{0})=\mathbb{E}_{\eta_{0}}[r(\alpha_{1}%
,\alpha_{2})],$ under the following mild condition.

\begin{assumption}
\label{UD} (i) Let the set $\mathcal{T}\subseteq\lbrack0,\infty)$ be a convex
set with a non-empty interior; (ii) the moment function $r$ is locally bounded.
\end{assumption}

\begin{proposition}
\label{PropABS}Under Assumption \ref{UD}, if $\phi(\eta_{0})=\mathbb{E}%
_{\eta_{0}}[r(\alpha_{1},\alpha_{2})]$ is regularly identified, then%
\[
r(\cdot)\in\left\{  b(\alpha_{1},\alpha_{2})\in L_{2}^{0}(\eta_{0}%
):b(\alpha_{1},\alpha_{2})=C_{1}+C_{2}\alpha_{2}^{2}e^{2\alpha_{1}\alpha_{2}%
}h(\alpha_{1}^{2},\alpha_{2}^{2})\right\}  ,
\]
for constants $C_{1}$ and $C_{2}$ and a continuous function $h(u,v)$ defined
on $(0,\infty)^{2}$ that, if $\mathcal{T}$ is bounded, is an infinite number
of times differentiable at $u\in(0,\infty),$ for all $v\in(0,\infty).$
\end{proposition}

\noindent For the purpose of proving an infinite efficiency bound for CDFs and
quantiles only the continuity part of Proposition \ref{PropABS} suffices.
Thus, an implication of Proposition \ref{PropABS} is that the CDF of UH at the
fixed point $(\alpha_{10},\alpha_{20}),$ i.e. $\phi(\eta_{0})=\mathbb{E}%
\left[  1(\alpha_{1}\leq\alpha_{10})1(\alpha_{2}\leq\alpha_{20})\right]  ,$ is
not regularly identified because $r_{\phi}(\alpha_{1},\alpha_{2})=1(\alpha
_{1}\leq\alpha_{10})1(\alpha_{2}\leq\alpha_{20})$ is not continuous when
$(\alpha_{10},\alpha_{20})$ is in the interior of the support of $\eta_{0}$.

\begin{corollary}
\label{CorcdfSU}Under Assumption \ref{UD}(i), the CDFs and quantiles of UH in
the model (\ref{densityEx1}) are not regularly identified.
\end{corollary}

\section{Random Coefficient Models}

\label{RC}

Random coefficient models have long been used in economics to model
nonparametric UH. There is by now an extensive literature on nonparametric
identification of UH in these models, see, e.g., Masten (2017) and references
therein. In this paper we focus on establishing irregular identification of
CDFs and quantiles of the distributions of RC. To the best of our knowledge,
this is the first paper to do so in this generality.

A general class of random coefficient models, including nonlinear models, is
given by%
\begin{equation}
Y_{i}=m\left(  X_{i},\alpha_{i}\right)  ,\label{GRC}%
\end{equation}
where $Z_{i}=(Y_{i},X_{i})$ are observed, but $\alpha_{i}$ is unobserved and
independent of $X_{i}$ with support $\mathcal{A}$. Assume $m:\mathcal{X}%
\times\mathcal{\mathcal{A}}\rightarrow\mathbb{R}^{r}$ is a measurable map,
where $\mathcal{X}$ is the support of $X$. The functional form of $m$ is
known, and the nonparametric part is given by the distribution of $\alpha
_{i}.$ The assumptions of known $m$ and the independence of $\alpha_{i}$ and
$X_{i}$ are relaxed below. The density of the data is%
\[
f_{\eta_{0}}(y,x)=\int_{\mathcal{A}}1\left(  y=m(x,\alpha)\right)  d\eta
_{0}(\alpha),
\]
where $1(A)$ denotes the indicator function of the event $A$. In this setting,
the dominating measure $\mu$ is defined on $\mathcal{Z}=\mathcal{Y}%
\times\mathcal{X}$ as $\mu\left(  B_{1}\times B_{2}\right)  =\nu_{Y}\left(
B_{1}\right)  \nu_{X}(B_{2}),$ where $B_{1}$ and $B_{2}$ are Borel sets of
$\mathcal{Y}$ and $\mathcal{X}$, respectively, $\nu_{Y}$ is either the
counting measure for discrete outcomes or the Lebesgue measure $\lambda
(\cdot)$ for continuous outcomes, and $\nu_{X}(\cdot)$ is the probability
measure for $X.$ The main challenge we face with RC models is that
$f_{z/\alpha}(z)=1\left(  y=m(x,\alpha)\right)  $ is not continuous, and thus
the previous results need to be generalized. The generalization is
non-trivial, particularly for continuous outcomes, and in some cases it
requires delicate technical work. We consider first the binary choice RC
model. Section \ref{NRC} in the Appendix contains some generic results for
nonlinear RC, as well as discussion on some RC models for which our
conclusions do not hold.

\subsection{Binary Choice Random Coefficient}

\label{ExampleRC}

The binary choice random coefficient model is given by
\[
Y_{i}=1\left(  X_{i}^{\prime}\alpha_{i}\geq0\right)  ,
\]
where we observe $Z_{i}=(Y_{i},X_{i})$ but $\alpha_{i}$ is unobservable. The
random vector $\alpha_{i}$ is independent of $X_{i},$ normalized to
$\left\vert \alpha_{i}\right\vert =1$ and satisfies $\mathbb{P}\left(
\alpha_{i}=0\right)  =0$. As in the existing literature, we assume $\eta_{0}$
is absolutely continuous wrt the uniform spherical measure $\sigma\left(
\cdot\right)  $ in $\mathbb{S}^{d_{\alpha}-1},$ where $\mathbb{S}^{d_{\alpha
}-1}=\{b\in\mathbb{R}^{d_{\alpha}}:\left\vert b\right\vert =1\}$ denotes the
unit sphere in $\mathbb{R}^{d_{\alpha}}.$ The density of the data for a
positive outcome (i.e. the choice probability function) is given by%
\begin{equation}
f_{\eta_{0}}(x)=\int_{\mathbb{S}^{d_{\alpha}-1}}1\left(  x^{\prime}%
s\geq0\right)  d\eta_{0}(s).\label{CP}%
\end{equation}
Ichimura and Thompson (1998) and Gautier and Kitamura (2013) found sufficient
conditions for nonparametric identification of $\eta_{0},$ but they did not
investigate whether identification was regular or irregular, which is the
focus here.

By (\ref{CP}) and Lemma \ref{Regu} a necessary condition for regular
identification of $\phi(\eta_{0})=\mathbb{E}_{\eta_{0}}[r(\alpha)]$ under
nonparametric UH is
\begin{equation}
r(\alpha)-\phi(\eta_{0})=\int1\left(  x^{\prime}\alpha\geq0\right)
s(1,x)dv_{X}(x),\label{regBRC}%
\end{equation}
for some $s\in L_{2}^{0}.$ The following result provides necessary conditions
for regular identification. Write $\alpha=(\alpha_{1},\alpha_{2}^{\prime
})^{\prime}$.

\begin{proposition}
\label{PropBRC}If the distribution of $X/\left\vert X\right\vert $ is
absolutely continuous, then $r(\cdot)$ in (\ref{regBRC}) must be uniformly
continuous on $\mathbb{S}^{d_{\alpha}-1}.$ If $X=(1,\tilde{X})\ $and
$\alpha_{2}^{\prime}\tilde{X}$ is absolutely continuous, then $r(\alpha
_{1},\alpha_{2})$ is an absolutely continuous function of $\alpha_{1}$.
\end{proposition}

\noindent An implication of this proposition is that functionals such as the
CDF and quantiles of random coefficients are not regularly identified in the
binary RC model. To the best of our knowledge, this result is new in the literature.

\begin{corollary}
\label{CorcdfBRC}Under the conditions of Proposition \ref{PropBRC}, the CDFs
and quantiles of UH in the binary RC model are not regularly identified.
\end{corollary}

\subsection{Linear Random Coefficient}

The linear RC model has a long history in econometrics, see, e.g., Hildreth
and Huock (1968) and Swamy (1970). This model is given by
\[
Y_{i}=X_{i}^{\prime}\alpha_{i},
\]
where we observe a $d_{z}-$dimensional vector $Z_{i}=(Y_{i},X_{i}),$ but
$\alpha_{i}$ is unobservable and independent of $X_{i}$. The dimension of
$X_{i}$ and $\alpha_{i}$ is $d_{\alpha},$ so $d_{z}=d_{\alpha}+1.$ Like in
Hoderlein, Klemel\"{a} and Mammen (2010), we normalize $X_{i}$ so that
$\left\vert X_{i}\right\vert =1.$ The density of the data is
\begin{equation}
f_{\eta_{0}}(z)=\int_{\mathbb{R}^{d_{\alpha}}}1\left(  y=x^{\prime}%
\alpha\right)  d\eta_{0}(\alpha). \label{densityLRC}%
\end{equation}
Nonparametric identification and estimation of $\eta_{0}$ has been studied by
Beran and Hall (1992), Beran, Feuerverger and Hall (1996), and Hoderlein,
Klemel\"{a} and Mammen (2010), among others. These authors exploit the
relation between (\ref{densityLRC}) and the Radon transform. In this paper we
study necessary conditions for regular identification of $\phi(\eta
_{0})=\mathbb{E}_{\eta_{0}}[r(\alpha)],$ for a measurable function $r\left(
\cdot\right)  $ with $\mathbb{E}_{\eta_{0}}[r^{2}(\alpha)]<\infty$, and
regular identification of quantiles of the components of $\alpha.$

By Lemma \ref{Regu} a necessary condition for regular identification of
$\phi(\eta_{0})=\mathbb{E}_{\eta_{0}}[r(\alpha)]$ under nonparametric UH is
\begin{equation}
r(\alpha)-\phi(\eta_{0})=\int s(x^{\prime}\alpha,x)dv_{X}(x), \label{RLRC}%
\end{equation}
for some $s\in L_{2}^{0}.$ Under suitable conditions scores in the tangent
space $\mathcal{S}=\{s\in L_{2}^{0}:s(z)=\mathbb{E}\left[  \left.
b(\alpha)\right\vert Z\right]  $ for some $b\in T(\eta_{0})\}$ are continuous,
but providing conditions under which elements of the \textit{closure} of
$\mathcal{S}$ are continuous is much harder. In fact, without additional
restrictions elements in the closure of $\mathcal{S}$ can be potentially very
discontinuous. We shall provide regularity conditions below that guarantee
that any element of the closure of $\mathcal{S}$ can be written as
\[
s(z)=\frac{g(z)}{f_{\eta_{0}}(z)},
\]
where $g(z)$ has an squared integrable weak derivative with respect to the
first argument $y.$ As we show below, this last condition will be instrumental
for checking the sufficient conditions for the dominated convergence theorem
in Lemma \ref{Main}.

Let $\eta_{0,x}$ denote the Lebesgue density of $x^{\prime}\alpha$ when
$\alpha$ has distribution $\eta_{0}.$ The set $\eta_{0}T(\eta_{0})$ is defined
as $\eta_{0}T(\eta_{0}):=\{\eta_{0}b:b\in T(\eta_{0})\},$ while the definition
of a Sobolev space $H^{\rho_{0}}(\mathcal{A})$ is provided after
(\ref{Sobnorm}) in the Appendix.

\begin{assumption}
\label{ALRC} For $d_{\alpha}>1$ and $N$ as in Assumption \ref{smooth}: (i) the
distribution $\eta_{0}$ is bounded, has bounded support, with a corresponding
density $\eta_{0,x}$ that is continuous and satisfies $\inf_{\alpha\in N}%
\eta_{0,x}(x^{\prime}\alpha)\geq1/l(x)$ for a positive measurable function
$l(\cdot)$ such that $\mathbb{E}_{X}[l^{2}(X)]<\infty$; (ii) $X$ is absolutely
continuous with a bounded density $f_{X}(\cdot);$ (iii) $\eta_{0}T(\eta
_{0})\subseteq H^{\rho_{0}}(\mathcal{A}),$ where $\rho_{0}+(d_{\alpha
}-1)/2>2;$ (iv) $r$ belongs to the closure of $T(\eta_{0}).$
\end{assumption}

The bounded support of Assumption \ref{ALRC}(i) is often considered in the
literature, see, e.g., Hoderlein, Klemel\"{a} and Mammen (2010). If the
infinite efficiency bound holds in a model with bounded support of $\alpha$ it
also holds in the more general model where the support is unrestricted. A
sufficient condition for the continuity of $\eta_{0,x}$ is that the Fourier
transform of the density of $\eta_{0}$ is integrable, which was also assumed
in Hoderlein, Klemel\"{a} and Mammen (2010). Assumptions \ref{ALRC}(i-ii)
establish a link between the tails of $\eta_{0}$ and $f_{X}(\cdot).$
Assumption \ref{ALRC}(iii) imposes a mild smoothness condition on the tangent
space of UH. This assumption and Assumption \ref{ALRC}(iv) allow but do not
require nonparametric UH.

\begin{proposition}
\label{PropLRC}Under Assumption \ref{ALRC} and if $r\ $satisfies (\ref{RLRC}),
then it must be continuous on $N.$
\end{proposition}

\begin{corollary}
\label{CorcdfLRC}Under the conditions of Proposition \ref{PropLRC}, the CDFs
and quantiles of UH are not regularly identified in the linear RC model.
\end{corollary}

\subsection{Correlated Random Coefficients: AME}

\label{NonpRegular}

The independence assumption between regressors and UH rules out important
models and parameters in economics, such as the Average Marginal Effect (AME)
$\phi(\eta_{0})=\mathbb{E}_{\eta_{0}}\left[  \gamma_{i}\right]  $ and the
Proportion of individuals with a Positive AME (PPAME), $\phi(\eta
_{0})=\mathbb{E}_{\eta_{0}}\left[  1\left(  \gamma_{i}>0\right)  \right]  ,$
where $\gamma_{i}$ is the coefficient of an endogenous continuous variable in
a RC triangular system. We extend our previous results to these cases. We will
show that under nonparametric UH these important parameters are not regularly
identified. These results appear to be new in the literature under this
generality. For simplicity, we focus on a triangular model, but the same
arguments are applicable to a wide class of random coefficient models,
including simultaneous equation models, nonlinear models with endogeneity, or
variations of these models that include covariates, multiple endogenous
variables, and mixed random and non-random coefficients.

Consider the triangular model:%
\begin{equation}
Y_{1}=\gamma Y_{2}+U_{1},\qquad Y_{2}=\delta X+U_{2},\label{triangularRC}%
\end{equation}
where $\gamma,$ $U_{1},$ $\delta$ and $U_{2}$ are RC, and we observe
$Z=(Y_{1},Y_{2},X)^{\prime}.$ The variable $Y_{2}$ is a continuous treatment
variable, possibly endogenous, in the sense that $U_{1}$ and $U_{2}$ are
correlated, and $X$ is an instrument, independent of all the random
coefficients. Suppose, the researcher is interested in the AME $\phi(\eta
_{0})=\mathbb{E}_{\eta_{0}}\left[  \gamma\right]  $ or the PPAME $\phi
(\eta_{0})=\mathbb{E}_{\eta_{0}}\left[  1(\gamma>0)\right]  .$ We will provide
conditions under which both parameters have an infinite efficiency bound. To
see this, we obtain the reduced forms%
\begin{align*}
Y_{1} &  =\gamma\delta X+\gamma U_{2}+U_{1}\equiv\pi_{1}X+\pi_{0},\\
Y_{2} &  =\delta X+U_{2},
\end{align*}
which, with some abuse of notation, are jointly written as $Y=\alpha
_{0}+\alpha_{1}X,$ where $Y=(Y_{1},Y_{2})^{\prime},$ $\alpha=(\alpha
_{0},\alpha_{1}),$ $\alpha_{0}=(\pi_{0},U_{2})^{\prime}$ and $\alpha_{1}%
=(\pi_{1},\delta)^{\prime}.$ Proposition \ref{PropLRC} can then be applied to
the reduced form. Because the corresponding influence functions for the AME
and PPAME are $r_{AME}(\alpha)=\pi_{1}/\delta$ and $r_{PPAME}(\alpha
)=1(\pi_{1}>0)1(\delta>0)+1(\pi_{1}<0)1(\delta<0),$ respectively, and they are
discontinuous functions of $\alpha_{1}=(\pi_{1},\delta)^{\prime}$,
non-regularity follows from Proposition \ref{PropLRC}. Consider the following
assumption. Let $N$ be an open set in the interior of $\mathcal{A},$ the
support of the reduced form random coefficient $\alpha$.

\begin{assumption}
\label{ACRC} (i) Assumption \ref{ALRC} holds with the reduced form
$Y=\alpha_{0}+\alpha_{1}X$; (ii) $X$ independent of the random coefficients
$(\gamma,U_{1},\delta,U_{2})$; (iii) $(p_{0},u_{2},0,d_{0})\in N$ for some
$(p_{0},u_{2},d_{0});$ (iv) $(p_{0},u_{2},p_{1},0)\in N$ for some
$(p_{0},u_{2},p_{1})$.
\end{assumption}

\begin{proposition}
\label{ATE} Suppose (\ref{triangularRC}) and Assumption \ref{ACRC}(i-ii)
holds. If in addition Assumption \ref{ACRC}(iii) or Assumption \ref{ACRC}(iv)
holds, then the PPAME is not regularly identified. If Assumption
\ref{ACRC}(iv) holds and $\mathbb{E}\left[  \gamma^{2}\right]  <\infty,$ then
the AME is not regularly identified.
\end{proposition}

\noindent Proposition \ref{ATE} proves non-regularity for the AME and the
PPAME. The condition $\mathbb{E}\left[  \gamma^{2}\right]  <\infty$ ensures
that the AME is a continuous functional in $L_{2}(\eta_{0})$. If
$f_{\delta^{2}}$ denotes the (Lebesgue) density of $\delta^{2}$ and
$h(u)=\mathbb{E}\left[  \left.  \pi_{1}^{2}\right\vert \delta^{2}=u\right]
f_{\delta^{2}}(u),$ then a sufficient condition for $\mathbb{E}\left[
\gamma^{2}\right]  <\infty$ is $\lim_{u\rightarrow0^{+}}h(u)/u^{\rho}<\infty$
for some $\rho>0$ and $\mathbb{E}\left[  \pi_{1}^{2}\right]  <\infty;$ see
Khuri and Casella (2002, pg. 45).

Intuitively, non-regularity of the AME comes from the presence of a set of
individuals with near-zero first-stage effects (Assumption \ref{ACRC}(iv)),
although $\mathbb{P}\left(  \delta=0\right)  =0$. When the instrument
satisfies a monotonicity restriction, in the sense that $\mathbb{P}%
(\delta>0)=1$ or $\mathbb{P}(\delta<0)=1,$ then regular identification of the
AME might be possible. Indeed, Heckman and Vytlacil (1998) and Wooldridge
(1997, 2003, 2008) show that with homogenous first-stage effects regular
estimation by IV methods holds. Masten (2017, Proposition 4) gives conditions
for nonparametric identification of the distribution of $\gamma,$ but he did
not discuss efficiency bounds for the AME or the PPAME under his conditions.
Khan and Tamer (2010) and Graham and Powell (2012) show irregularity of the
AME in different models where $\mathbb{E}\left[  \gamma^{2}\right]  =\infty$.
We show irregularity of the AME in a setting where $\mathbb{E}\left[
\gamma^{2}\right]  <\infty.$ See also Florens et al. (2008), Masten and
Torgovitsky (2016), and the extensive literature following the seminal
contributions by Imbens and Angrist (1994) and Heckman and Vytlacil (2005) for
identification results on conditional and weighted AME or their discrete versions.

The PPAME is non-regular under more general conditions than the AME, because
it has a discontinuous influence function under more general conditions than
that of the AME. Heckman, Smith and Clements (1997) provide bounds for the
analog to PPAME in the binary treatment case, and identification when gains
are not anticipated at the time of the program. The irregularity of the PPAME
also follows from a more general principle that we describe in the next
section: if irregularity holds in a model with exogenous effects, it also
holds in the model with endogenous effects.

\section{Extension to Semiparametric Models}

\label{SemiparametricModels}

This section extends our results to semiparametric models. The main point is
as follows, if a functional is non-regularly identified in a model, it will be
non-regularly identified in a larger model that nests the original model as a
special case. Information can only decrease (or remain the same) when we know
less. This basic observation has important implications, and it widens
substantially the applicability of our results as illustrated with the Mixed
Logit model here and with further examples in the Appendix.

\subsection{The Mixed Logit Model}

\label{mixedlogit}

Consider first a conditional semiparametric mixture model with density%
\[
f_{\eta_{0},\theta_{0}}(y,x)=\int f_{y/x,\alpha}(y;\theta_{0})d\eta_{0}%
(\alpha),
\]
where $\theta_{0}$ is an additional unknown parameter, finite or
infinite-dimensional. The basic idea here is that irregularity of $\phi
(\eta_{0})=\mathbb{E}_{\eta_{0}}[r(\alpha)]$ in the model where $\theta_{0}$
is known implies irregularity in the model where $\theta_{0}$ is unknown.

We illustrate our point with the random coefficients Logit model, also known
as the Mixed Logit---one of the most commonly used models in applied choice
analysis. Fox, Kim, Ryan and Bajari (2012) have recently shown nonparametric
identification for the semiparametric Mixed Logit model. Here, we show that
the identification of the CDF and quantiles of the distribution of RC is
necessarily irregular when UH is nonparametric. The CDF and quantiles of this
distribution are important parameters in applications of discrete choice.

The data $Z_{i}=(Y_{i},X_{i})$ is a random sample from the density (wrt $\mu$
below),%
\[
f_{\lambda_{0}}(y,x)=\int f_{y/x,\alpha}(y;\theta_{0})d\eta_{0}(\alpha),
\]
where $\lambda_{0}=(\theta_{0},\eta_{0})\in\Theta\times H,$ $\theta
_{0}=(\theta_{01},...,\theta_{0J})^{\prime},$
\[
f_{y/x,\alpha}(y;\theta_{0})=\frac{\exp\left(  \theta_{0y}+x_{y}^{\prime
}\alpha\right)  }{1+\sum_{j=1}^{J}\exp\left(  \theta_{0j}+x_{j}^{\prime}%
\alpha\right)  },
\]
for $x=(x_{0},x_{1},...,x_{J})\in\mathcal{X}$ and $y\in\mathcal{Y}=\left\{
0,1,...,J\right\}  $. The consumer can choose between $j=1,...,J,$ $J<\infty,$
mutually exclusive inside goods and one outside good $(y=0).$ The utility for
the inside good is normalized so that $\theta_{00}=0$ and $x_{0}=0.$ The
random coefficients $\alpha$ are independent of the regressors $X,$ and have a
distribution $\eta_{0}.$ The main result below also applies to the correlated
random coefficient case$.$ In fact, non-regular identification for CDFs and
quantiles is proved even when $\theta_{0}$ is known. This will imply
non-regularity when $\theta_{0}$ is unknown and/or when random coefficients
are dependent of the characteristics.

The measure $\mu$ is defined on $\mathcal{Z}=\mathcal{Y}\times\mathcal{X}$ as
$\mu\left(  B_{1}\times B_{2}\right)  =\tau\left(  B_{1}\right)  \nu_{X}%
(B_{2}),$ where $B_{1}\subset\mathcal{Y}$, $B_{2}$ is a Borel set of
$\mathcal{X}$, $\tau(\cdot)$ is the counting measure and $\nu_{X}(\cdot)$ is
the probability measure for $X.$ The vector $\alpha$ and covariates $x_{y}$
are $K-$dimensional. The parameter space $\Theta$ is an open set of
$\mathbb{R}^{J}.$ The set $H$ consists of measurable functions $\eta
:\mathbb{R}^{K}\rightarrow\mathbb{R}$ whose support $\mathcal{A}$ has a
non-empty interior and $\int_{\mathcal{A}}d\eta(\alpha)=1$.

Applying the necessary condition for regular identification to a continuous
linear functional $\phi(\eta)\in\mathbb{R}$ with influence function $r_{\phi}$
in the model where $\theta_{0}$ is known, it must be true that for some $s\in
L_{2},$
\begin{equation}
r_{\phi}(\alpha)-\phi(\eta_{0})=\int f_{y/x,\alpha}(y;\theta_{0}%
)s(y,x)d\mu(y,x).\label{integral2}%
\end{equation}
It is straightforward to show that the right hand side in (\ref{integral2}) is
continuous in $\alpha$ in the interior of its support. In fact, more is true
in general: it is an analytic function of $\alpha$ (a function that is
infinitely differentiable with a convergent power series expansion). But
continuity suffices for proving the non-regularity of CDFs and quantiles of
$\eta_{0}.$ This follows without computing least favorable distributions and
efficiency bounds, simply by dominated convergence. We gather the proof here
to illustrate the simplicity of our method of proof.

\begin{proposition}
\label{zeroinfologit} $r_{\phi}\ $in (\ref{integral2}) is continuous in the
interior of $\mathcal{A}$.
\end{proposition}

\noindent\textbf{Proof of Proposition \ref{zeroinfologit}}: Write%
\[
\int f_{y/x,\alpha}(y;\theta_{0})s(y,x)d\mu(y,x)=\sum_{j=0}^{J}\int
f_{y/x,\alpha}(j;\theta_{0})s(j,x)v_{X}(dx).
\]
Each of the summands in the last expression is continuous in $\alpha$ in the
interior of its support, by continuity and boundedness of $f_{y/x,\alpha
}(j;\theta_{0})$ and the dominated convergence theorem. $\blacksquare$

Proposition \ref{zeroinfologit} implies that identification of the CDF and
quantiles of the distribution of $\eta_{0}$ under the conditions specified in
Fox et al. (2012) must be irregular. Bajari, Fox and Ryan (2007) propose a
simple estimator of the CDF of $\eta_{0}$, and Fox, Kim and Yang (2016) show
its consistency (in the weak topology) and obtain its rates of convergence$.$
Proposition \ref{zeroinfologit} implies that the estimator in Fox et al.
(2016), or any other estimator for that matter, cannot achieve regular
parametric rates of convergence. The lack of regularity is not evident from
the rates established in Fox et al. (2016). Let $F_{0}$ be the CDF pertaining
to $\eta_{0}$ and $\widehat{F}_{\eta}$ the \textquotedblleft fixed
grid\textquotedblright\ estimator of Bajari et al. (2007), Fox et al. (2011)
and Fox et al. (2016) based on $D$ grid points $(D\equiv D(n),$ where $n$ is
the sample size). The order of the bias established in Fox et al. (2016) is
$D^{-\bar{s}/K}$ where $\bar{s}$ is the smoothness of the mapping
$\alpha\rightarrow f_{y/x,\alpha}$ (here $\bar{s}=\infty).$ This suggests that
parametric rates might be attainable, but our results show that this is not
possible (at least in a local uniform sense). The order of the variance for
$\widehat{F}_{\eta}$ is inversely related to the minimum eigenvalue of the
$D\times D$ matrix $\Psi_{D}$ with $(d_{1},d_{2})-th$ element, $1\leq
d_{1},d_{2}\leq D,$ given by
\begin{equation}
\mathbb{E}\left[  g^{\prime}(X,\alpha_{d_{1}})g(X,\alpha_{d_{2}})\right]
,\label{design}%
\end{equation}
where $g(x,\alpha_{d})=(f_{y/x,\alpha_{d}}(0;\theta_{0}),...,f_{y/x,\alpha
_{d}}(J;\theta_{0}))^{\prime}$ are conditional choice probabilities when UH is
evaluated at the $d-th$ grid point $\alpha_{d},$ $d=1,...,D.$ This minimum
eigenvalue quantifies the level of multicollinearity in the least squares
regression of Fox et al. (2016), and we conjecture that given the high
smoothness of the mapping $\alpha\rightarrow f_{y/x,\alpha}$ this term will go
to zero exponentially fast, so it will be the main determinant in the (slow)
rate of convergence of $\widehat{F}_{\eta}$. A detailed theoretical analysis
of this issue is beyond the scope of this paper, but see the discussion in the
next section and the Monte Carlo simulations below, which support these claims.

\section{Regularization}

\label{Reg}

The previous examples show that regular identification of CDFs and quantiles
of UH in the models considered may require restricting the nature of
heterogeneity. In this section we investigate how common approaches considered
in the literature address the lack of regularity of these functionals.
Additionally, we provide a necessary condition for CDFs and quantiles to be
regularly identified when UH is semiparametric and a discussion on how
smoothness of $\alpha\rightarrow f_{z/\alpha}$ translates into a
multicollinearity problem for sieve and related estimators.

Our first observation is derived from the main idea in the previous section:
functional form assumptions that restrict the conditional likelihood may not
help with the irregular identification of CDFs and quantiles if still the
mapping $\alpha\rightarrow f_{z/\alpha}$ is smooth, while UH is nonparametric.
For example, knowing the finite dimensional parameters of a semiparametric
mixture, knowing the functional forms of the idiosyncratic error terms in
Kotlarski's lemma, or knowing the functional form of the baseline hazard in
the mixed proportional hazard model do not help in restoring regular
identification of CDFs and quantiles of UH when UH\ is nonparametric.

We discuss how restrictions on UH translate into regularity of functionals of
UH. Denote by $\overline{T(\eta_{0})}$ the mean squared closure of $T(\eta
_{0})$ in $L_{2}(\eta_{0}).$ That UH is not nonparametric formally means that
$\overline{T(\eta_{0})}$ is a strict subset of $L_{2}^{0}(\eta_{0}).$ The
extension of the necessary condition for regular identification of $\phi
(\eta_{0})=\mathbb{E}_{\eta_{0}}[r(\alpha)],$ for a measurable function
$r\left(  \cdot\right)  $ with $\mathbb{E}_{\eta_{0}}[r^{2}(\alpha)]<\infty,$
is given in the following lemma. Let $\Pi_{\overline{V}}$ denote the
orthogonal projection operator onto $\overline{V},$ where $\overline{V}$
denotes the closure of $V$ in the norm topology.

\begin{lemma}
\label{ReguSem}The necessary condition for regular identification of
$\phi(\eta_{0})=\mathbb{E}_{\eta_{0}}[r(\alpha)]$ is
\begin{equation}
\Pi_{\overline{T(\eta_{0})}}r(\alpha)=\Pi_{\overline{T(\eta_{0})}}%
\mathbb{E}\left[  \left.  s(Z)\right\vert \alpha\right]  ,\text{ for some
}s\in L_{2}^{0}. \label{regsem}%
\end{equation}

\end{lemma}

\noindent The mismatch in smoothness between $r(\alpha)$ and $\mathbb{E}%
\left[  \left.  s(Z)\right\vert \alpha\right]  ,$ which was the source of
irregularity in the examples studied, may now be restored by the projection
onto $\overline{T(\eta_{0})}.$ We briefly discuss how different restrictions
on UH translate into regularity of CDFs and quantiles in view of this general characterization.

A popular approach in practice is to consider a parametric distribution for
the UH. A leading example of parametric model is a finite mixture with known
and finite support points. Parametric heterogeneity leads to a finite
dimensional tangent space $T(\eta_{0}),$ which is then closed $T(\eta
_{0})=\overline{T(\eta_{0})},$ and which is generated by the scores of the
specified distribution. Denote by $l_{\eta}$ the score of UH, i.e.
$\overline{T(\eta_{0})}=T(\eta_{0})=span(l_{\eta}),$ assume $\mathbb{E}%
_{\eta_{0}}\left[  l_{\eta}(\alpha)l_{\eta}^{\prime}(\alpha)\right]  $ is
non-singular, and define the projected score $s_{0}(Z)=\mathbb{E}\left[
\left.  l_{\eta}(\alpha)\right\vert Z\right]  .$ Then, simple algebra shows
that a solution to (\ref{regsem}) in $s$ is given by $s_{r}$ defined by%
\[
s_{r}(Z)=\lambda_{r}^{\prime}s_{0}(Z),
\]
where $\lambda_{r}$ is a solution to
\begin{equation}
\mathbb{E}\left[  s_{0}(Z)s_{0}^{\prime}(Z)\right]  \lambda_{r}=\mathbb{E}%
\left[  r(\alpha)l_{\eta}^{\prime}(\alpha)\right]  . \label{system}%
\end{equation}
If the Fisher information for $\eta_{0}$ is positive, which means
$\mathbb{E}\left[  s_{0}(Z)s_{0}^{\prime}(Z)\right]  $ is non-singular, then
there is a unique solution $\lambda_{r}$ of (\ref{system}), and $\phi(\eta
_{0})$ is regularly identified. More generally, $\phi(\eta_{0})$ may be
regularly identified even when $\eta_{0}$ is not, and this corresponds to the
system in (\ref{system}) having some solution in $\lambda_{r}.$ The drawback
of the parametric approach is the high misspecification risk, which can be
quantified by the dimension and form of the model's tangent space. If the
dimension of $T(\eta_{0})$ is $D,$ then the tangent space of the model is at
most $D-$dimensional and given by $\mathcal{S}:=\{s\in L_{2}^{0}%
:s(z)=\lambda^{\prime}s_{0}(z)$ for some $\lambda\in\mathbb{R}^{D}\}.$
Estimators for functionals of UH will be in general inconsistent when the
model is misspecified.

As usual, a semiparametric approach is more robust to misspecification. In
Lemma \ref{ReguSem} we have derived the necessary condition for regular
identification of moments when UH is semiparametric, so $\overline{T(\eta
_{0})}$ is a strict subset of $L_{2}^{0}(\eta_{0})\ $of infinite dimension.
Examples of semiparametric models include finite mixtures with unknown support
points and sieve methods with incomplete sieve basis. Existing rate results
for finite mixtures with unknown support points suggest irregularity of the
CDFs in general (see, e.g., Chen 1995 and Heinrich and Kahn 2018), although we
are not aware of any paper investigating semiparametric efficiency bounds for
finite mixtures with unknown support points. We recognize that, although the
sufficient condition for semiparametric restrictions in Lemma \ref{ReguSem} is
general, it may be hard to find primitive conditions for it, as computing the
closure of $T(\eta_{0})$ and the projections onto it may not be
straightforward in applications.

As a practical approach, we recommend a sieve method where the span of
$\{l_{\eta}(\alpha)\}$ increases with the sample size, i.e. $D\rightarrow
\infty$ as $n\rightarrow\infty$. Without loss of generality normalize
$l_{\eta}$ so that $\mathbb{E}_{\eta_{0}}\left[  l_{\eta}(\alpha)l_{\eta
}^{\prime}(\alpha)\right]  $ is the identity matrix. A key quantity for sieve
estimation is the minimum eigenvalue of the Fisher information matrix
$\mathbb{E}\left[  s_{0}(Z)s_{0}^{\prime}(Z)\right]  ,$ denoted by $\xi_{\min
}\equiv\xi_{\min}(D);$ see Fox, Kim and Yang (2016) and (\ref{system}). We
provide a useful bound for $\xi_{\min}.$ To that end, we assume the score
operator $Ab=\mathbb{E}\left[  \left.  b(\alpha)\right\vert Z\right]  $ from
$L_{2}(\eta_{0})$ to $L_{2}$ is compact. A well known sufficient condition for
this is%
\begin{equation}
\int\frac{f_{z/\alpha}^{2}(z)}{f_{\eta_{0}}(z)}d\eta_{0}(\alpha)d\mu
(z)<\infty.\label{compact}%
\end{equation}
Under this condition, $A$ has a sequence of singular values $\{\mu_{d}%
\}_{d=1}^{\infty}$ (see Engl, Hanke and Nuebauer, 1996). Then, the following
bound follows essentially from Blundell, Chen and Kristensen (2007, Lemma 1).

\begin{lemma}
\label{Bound}If (\ref{compact}) holds, then $\xi_{\min}(D)\leq\mu_{D}^{2}.$
\end{lemma}

Since $\mu_{D}\rightarrow0$ as $D\rightarrow\infty,$ Lemma \ref{Bound} implies
that also $\xi_{\min}(D)\rightarrow0$. This is the multicollinearity problem
mentioned above. Furthermore, the score operator $A$ is an integral operator
with kernel $K(z,\alpha)=f_{z/\alpha}(z)/f_{\eta_{0}}(z),$ and it is well
known that the smoother the mapping $\alpha\rightarrow K(z,\alpha),$ the
faster the singular values $\mu_{D}$ go to zero. In particular, for analytical
kernels the singular values decay exponentially fast to zero (Hille and
Tamarkin 1931). The minimum eigenvalue $\xi_{\min}(D)$ is also closely related
to the sieve measure of ill-posedness $\tau_{D}$ proposed in econometrics (see
Chen 2007 and Blundell, Chen and Kristensen 2007) through the relation%
\[
\tau_{D}^{2}=\frac{1}{\xi_{\min}(D)}.
\]
Prior to this paper, Blundell, Chen and Kristensen (2007, Lemma 1) obtained
the bound $\tau_{D}\geq1/\mu_{D}$ in a nonparametric IV setting. Thus, the
modest contribution here is the interpretation in terms of the minimum
eigenvalue of the Fisher information matrix. For applications of sieve
estimators along this line and the important role of $\tau_{D}$ (or $\xi
_{\min}(D)$) see, e.g., Chen (2007), Bajari, Fox and Ryan (2007), Hu and
Schennach (2008), Bester and Hansen (2007), Chen and Liao (2014), Fox, Kim and
Yang (2016) and references therein. Next section investigates the finite
sample performance of the sieve \textquotedblleft fixed grid\textquotedblright%
\ method of Fox, Kim and Yang (2016) and a regularized version to reduce the
variance of estimates of the CDFs and quantiles of UH.

\section{Monte Carlo}

\label{MC}

This section illustrates some of the theoretical ideas in a Monte Carlo study
on the Mixed Logit model. Specifically, we consider the \textquotedblleft
fixed grid\textquotedblright\ nonparametric estimator of Bajari et al. (2007)
and Fox et al. (2016), and evaluate the performance of this estimator for
estimating the CDF and quantiles of UH.\footnote{We thank Jeremy Fox for
sharing the Matlab code to implement their estimator.} We also provide a
variant of this estimator that performs a Singular Value Decomposition (SVD)
of the resulting design matrix to reduce the variance of the estimator. To
introduce the estimator, consider a discrete approximation of the distribution
of UH of the form%
\begin{equation}
\eta_{0}(\alpha)\approx\sum_{d=1}^{D}\theta_{d}\delta_{\alpha_{d}}%
(\alpha),\label{appr}%
\end{equation}
where $\theta_{d}$ are probabilities, adding up to one, over a finite support
$\{\alpha_{d}\}_{d=1}^{D}$ of size $D$ in $\mathcal{A}.$ In Fox et al. (2016)
$D,$ and thus the discrete support, is allowed to increase with the sample
size $n$. Define $Y_{i,j}$ as the binary choice equals 1 whenever individual
$i^{\prime}s$ choice is $j,$ and zero otherwise. Define the regression error
term $\varepsilon_{i,j}=Y_{i,j}-f_{\eta_{0}}(j,X_{i}).$ The least squares
estimator uses the regression equation
\[
Y_{i,j}=\int f_{y/X_{i},\alpha}(j)d\eta_{0}(\alpha)+\varepsilon_{i,j},
\]
with the approximation in (\ref{appr}) to obtain the approximated linear
regression model%
\[
Y_{i,j}\approx\sum_{d=1}^{D}\theta_{d}f_{y/X_{i},\alpha_{d}}(j)+\varepsilon
_{i,j}.
\]
Fox et al. (2016) proposes running a regression of $Y_{i,j}$ on the regressors
$Z_{i,j}^{d}:=f_{y/X_{i},\alpha_{d}}(j)$ subject to the constrains on the
probabilities $\theta_{d},$ i.e.%
\begin{equation}
\widehat{\theta}=\arg\min_{\theta\in\Delta_{d}}\frac{1}{nJ}\sum_{i=1}^{n}%
\sum_{j=0}^{J}\left(  Y_{i,j}-\sum_{d=1}^{D}\theta_{d}Z_{i,j}^{d}\right)
^{2},\label{LSEr}%
\end{equation}
where $\theta=(\theta_{1},...,\theta_{D})^{\prime}\in\Delta_{d}=\left\{
(p_{1},...,p_{D}):0\leq p_{d}\leq1\text{ and }\sum_{d=1}^{D}p_{d}=1\right\}
.$ The least squares problem in (\ref{LSEr}) is convex and can be efficiently
solved by standard routines (such as lsqlin in Matlab). The estimator of the
CDF of $\eta_{0}$ at $\alpha_{0}$ is then given by%
\begin{equation}
\widehat{F}_{\eta}(\alpha_{0})=\sum_{d=1}^{D}\widehat{\theta}_{d}1(\alpha
_{d}\leq\alpha_{0}),\label{CDFhat}%
\end{equation}
and from the CDF we define the quantile estimators as usual.

For simplicity of computation, in the Monte Carlo we apply this estimator to
the Mixed Logit model without fixed parameters, so
\[
f_{y/x,\alpha}(y)=\frac{\exp\left(  x_{y}^{\prime}\alpha\right)  }%
{1+\sum_{j=1}^{J}\exp\left(  x_{j}^{\prime}\alpha\right)  },
\]
for $x=(x_{0},x_{1},...,x_{J})\in\mathcal{X}$ and $y\in\mathcal{Y}=\left\{
0,1,...,J\right\}  $. Smoothness of mapping $\alpha\longrightarrow
f_{y/x,\alpha}$ translates into high correlation of the regressors
$Z_{i,j}^{d}$ when $D$ is large (for $d^{\prime}s$ corresponding to nearby
$\alpha_{d}^{\prime}s$), suggesting that methods that account for
multicollinearity may reduce the variances of the resulting estimators. We
suggest using the SVD of the design $nJ\times D$ matrix $\mathbf{Z}%
=(Z_{i,j}^{d}),$ by adding the linear constrain $V_{p-D}^{\prime}\theta=0$ to
(\ref{LSEr}), where $V_{p-D}=(v_{p-D},v_{p-D+1},...,v_{D})$ denotes the last
$p-D$ left singular vectors of $\mathbf{Z}$ (where as usual, they are ordered
according to the singular values from largest to smallest). This is the
classical Principal Component Regression adapted to the constrained case where
$\theta^{\prime}s$ are probabilities. The resulting estimator is
\[
\widetilde{\theta}=\arg\min_{\theta\in\Delta_{d},V_{p-D}^{\prime}\theta
=0}\frac{1}{nJ}\sum_{i=1}^{n}\sum_{j=0}^{J}\left(  Y_{i,j}-\sum_{d=1}%
^{D}\theta_{d}Z_{i,j}^{d}\right)  ^{2},
\]
which solves a convex problem and can be equally computed by routines such as
lsqlin in Matlab. Let $\widetilde{F}_{\eta}(\alpha_{0})=\sum_{d=1}%
^{D}\widetilde{\theta}_{d}1(\alpha_{d}\leq\alpha_{0})$ denote the
corresponding CDF estimator$.$ We compare below the performance of the
resulting CDFs and quantile estimators based on $\widehat{\theta}$ and
$\widetilde{\theta},$ respectively.

The Monte Carlo setting we consider is taken from a recent study by Heiss,
Hetzenecker and Osterhaus (2019). The data generating process we consider is
as follows. The number of products (not including outside good) is $J=3$. The
number of product characteristics is $K=2$. The characteristics are generated
as independent uniforms on $[0,1].$ The random coefficient distribution is a
mixture of two bivariate normal distributions with probability weights
$(1/2,1/2),$ means $(-2.2,-2.2)$ and $(1.3,1.3)$ and equal variances
$\Sigma_{1}=\Sigma_{2}=\Sigma$ given by%
\[
\Sigma=\left[
\begin{array}
[c]{cc}%
0.8 & 0.15\\
0.15 & 0.8
\end{array}
\right]  .
\]
To generate the grid $\{\alpha_{d}\}_{d=1}^{D}$ we use a Halton sequence with
points spread on $[-5,5]\times\lbrack-5,5].$ The fixed grid covers the support
of the true distribution with probability close to one. We consider different
values for the number of points in the grid $D\in\{25,100,500\}$ and sample
sizes $n\in\{100,500,1000\}.$ For computing $\widetilde{\theta}$ we set the
number of components $p$ to 5 throughout (we have investigated with values of
$p$ between 3 and 10 and obtain qualitatively similar results). We set $p$
deterministically in simulations to save time, but in practice we recommend
cross-validation to select $p$. The number of Monte Carlo simulations is
$M=500.$ To evaluate the performance of CDFs' estimators we compute the
integrated absolute bias%
\[
Bias(\widehat{F})=\frac{1}{ML}\sum_{m=1}^{M}\sum_{l=1}^{L}\left\vert
\widehat{F}_{\eta,m}(\alpha_{l})-F_{0}(\alpha_{l})\right\vert ,
\]
where $\{\alpha_{l}\}_{l=1}^{L}$ is an additional equally spaced grid over
$[-5,5]\times\lbrack-5,5]$ with $L=121,$ $\widehat{F}_{\eta,m}$ is the fixed
grid CDF estimator (cf. \ref{CDFhat}) for the $m-th$ Monte Carlo simulation,
and $F_{0}$ denotes the true CDF pertaining to $\eta_{0}$.

We also report the Root integrated Mean Squared Error defined as
\[
RMSE(\widehat{F})=\sqrt{\frac{1}{ML}\sum_{m=1}^{M}\sum_{l=1}^{L}\left(
\widehat{F}_{\eta,m}(\alpha_{l})-F_{0}(\alpha_{l})\right)  ^{2}}.
\]
The quantities $Bias(\widetilde{F})$ and $RMSE(\widetilde{F})$ are analogously defined.

Table 1 reports the bias and root mean squared errors for the CDFs estimators
$\widehat{F}$ and $\widetilde{F}.$ The first observation is that the bias is
small even for small sample sizes such as $n=100,$ and it does not depend much
on $D,$ which is consistent with our discussion in Section \ref{mixedlogit}.
The regularization causes $\widetilde{F}$ to have a slightly larger bias than
$\widehat{F}$ in some cases, although the difference is not substantial, and
for small samples the bias of $\widetilde{F}$ is even smaller. On the other
hand, the variance of $\widehat{F}$ is systematically larger than that of
$\widetilde{F},$ particularly for moderate and large values of $D,$ consistent
with our claims that the level of multicollinearity increases dramatically
with the number of points $D$. \bigskip

\begin{center}
\textbf{Table 1}. Bias and RMSE for CDFs in Mixed Logit%

\begin{tabular}
[c]{cccccc}\hline\hline
$n$ & $D$ & $Bias(\widehat{F})$ & $Bias(\widetilde{F})$ & $RMSE(\widehat{F})$
& $RMSE(\widetilde{F})$\\\hline
100 & 25 & 0.0781 & 0.0729 & 0.1791 & 0.1059\\
500 & 25 & 0.0663 & 0.0713 & 0.1380 & 0.0933\\
1000 & 25 & 0.0605 & 0.0708 & 0.1231 & 0.0904\\
100 & 100 & 0.0799 & 0.0682 & 0.1896 & 0.0999\\
500 & 100 & 0.0606 & 0.0639 & 0.1428 & 0.0855\\
1000 & 100 & 0.0511 & 0.0630 & 0.1284 & 0.0831\\
100 & 500 & 0.0784 & 0.0651 & 0.1906 & 0.0982\\
500 & 500 & 0.0541 & 0.0602 & 0.1452 & 0.0835\\
1000 & 500 & 0.0440 & 0.0592 & 0.1303 & 0.0805\\\hline
\end{tabular}

$M=500$ simulations.
\end{center}

Table 2 reports the RMSE for the medians of the marginal distributions of UH
(denoted by RMSEQ1 and RMSEQ2 for $\widehat{F}$ and RMSEQ1-PCR and RMSEQ2-PCR
for $\widetilde{F},$ respectively). Results for other quantile levels are
reported in the Appendix. We do not report the bias separately to save space,
but we note that the bias for quantiles is much larger than the bias for CDFs.
We observe substantial gains in terms of RMSE of the regularization by SVD,
with the benefits increasing with the number of grid points. Importantly, in
both cases, CDFs and quantiles, the reported results are consistent with much
slower rates of convergence than parametric, lending support on the infinite
efficiency bounds established in this paper. \bigskip

\begin{center}
\textbf{Table 2}. RMSE for Medians of Marginals of UH in the Mixed
Logit\newline%
\begin{tabular}
[c]{cccccc}\hline\hline
$n$ & $D$ & RMSEQ1 & RMSEQ1-PCR & RMSEQ2 & RMSEQ2-PCR\\\hline
100 & 25 & 1.6624 & 0.8061 & 1.4621 & 0.7085\\
500 & 25 & 0.8492 & 0.5232 & 0.8713 & 0.4155\\
1000 & 25 & 0.8008 & 0.4923 & 0.7386 & 0.3254\\
100 & 100 & 1.6084 & 0.6315 & 1.8392 & 0.6514\\
500 & 100 & 0.9411 & 0.2995 & 0.9409 & 0.2790\\
1000 & 100 & 0.8947 & 0.1874 & 0.8976 & 0.1832\\
100 & 500 & 1.6373 & 0.6360 & 1.6270 & 0.5974\\
500 & 500 & 1.0599 & 0.2710 & 0.9917 & 0.2639\\
1000 & 500 & 0.9374 & 0.1879 & 0.9669 & 0.1766\\\hline
\end{tabular}

$M=500$ simulations.
\end{center}

\section{Conclusions}

\label{Conclusions}

We have established irregular identification of CDFs and quantiles (or more
generally, functionals with discontinuous influence functions) of
nonparametric UH in some structural economic models. Example applications
include the structural model of unemployment with two spells in Alvarez et al.
(2015), the binary and linear RC models (possibly with correlated effects),
the AME in a triangular model with near zero first-stage effects, and the
distribution and quantiles of UH in the Mixed Logit model. These are only some
applications, but the results are applicable more widely. Further examples in
the Appendix include mixed proportional duration models, and measurement error
models with two measurements identified by means of Kotlarski's lemma.
Furthermore, as we discuss in the Appendix, we expect our approach to be
applicable to many situations where the so-called Information Operator (see
e.g. Begun, Hall, Huang and Wellner (1983)) is a smoothing operator.

The most appealing feature of our method of proof is its simplicity, relative
to alternative approaches that directly compute efficiency bounds, which are
particularly difficult to compute in the models we have studied. Instead, we
exploit some necessary smoothness conditions that the influence function of a
regularly identified functional must satisfy. The Mixed Logit example is
illustrative of the easiness in the application of our method of proof. In
contrast, directly computing the Fisher information and the efficiency bound
in this model is rather challenging (and were unknown prior to this paper).
The practical implications of the irregularity of CDFs and quantiles have been
investigated in a Monte Carlo study. We have found substantial benefits from
regularizing the fixed grid estimator of Bajari et al. (2007), Fox et al.
(2011) and Fox et al. (2016), without sacrificing much of its appealing
computational simplicity. Future research on the theoretical properties of
regularized estimators is guaranteed. \newpage

\section{Appendix A: Proofs of Main Results}

\label{Proofs}

\noindent\textbf{Proof of Lemma \ref{Regu}}: First, the functional $\eta
_{0}\rightarrow\phi(\eta_{0})=\mathbb{E}_{\eta_{0}}[r(\alpha)]$ is
differentiable with influence function%
\[
\chi(\alpha)=\Pi_{\overline{T(\eta_{0})}}r(\alpha),
\]
where $\Pi_{\overline{V}}$ denotes the orthogonal projection operator onto the
closure of $V,$ $\overline{V}.$ To see this, note that by linearity of
$\eta_{0}\rightarrow\phi(\eta_{0}),$ for all $b\in T(\eta_{0}),$%
\begin{align*}
\lim_{t\rightarrow0}\frac{\phi(\eta_{t})-\phi(\eta_{0})}{t} &  =\mathbb{E}%
_{\eta_{0}}[r(\alpha)b(\alpha)]\\
&  =\mathbb{E}_{\eta_{0}}[\left(  \Pi_{\overline{T(\eta_{0})}}r(\alpha
)\right)  b(\alpha)].
\end{align*}
Since UH is nonparametric $\Pi_{\overline{T(\eta_{0})}}r(\alpha)=r(\alpha
)-\phi(\eta_{0}).$ On the other hand, by Lemma 25.34 in van der Vaart (1998)
the adjoint of the score operator is given by
\[
A^{\ast}s=\mathbb{E}\left[  \left.  s(Z)\right\vert \alpha\right]
-\mathbb{E}\left[  s(Z)\right]  .
\]
The lemma then follows from Theorem 3.1 and Theorem 4.1 in van der Vaart
(1991), which establish that a necessary condition for positive Fisher
information for $\phi(\eta_{0})$ is
\[
r(\alpha)-\phi(\eta_{0})=\mathbb{E}\left[  \left.  s(Z)\right\vert
\alpha\right]  ,
\]
since $\mathbb{E}\left[  s(Z)\right]  =0$. $\blacksquare$\bigskip\

\noindent\textbf{Proof of Lemma \ref{Main}: }Let $\alpha_{n},\alpha\in N$ such
that $\alpha_{n}\rightarrow\alpha,$ and define $h_{n}(z)=s(z)f_{z/\alpha_{n}%
}(z).$ Note (i) implies $h_{n}(z)\rightarrow h(z):=s(z)f_{z/\alpha}(z)$
a.e-$\mu.$ Also, by the dominance condition, for a sufficiently large $n,$%
\[
\int\left\vert h_{n}(z)\right\vert d\mu(z)<\infty.
\]
We conclude by dominated convergence that%
\[
\int s(z)f_{z/\alpha_{n}}(z)d\mu(z)\rightarrow\int s(z)f_{z/\alpha}%
(z)d\mu(z).
\]
$\blacksquare$\bigskip

\noindent\textbf{Proof of Corollary \ref{Corcdf}:} By Lemma \ref{Main} if the
influence function of the functional is discontinuous then the functional is
not regularly identified. Since the indicator is not continuous, this proves
the lemma. $\blacksquare$\bigskip\

\noindent\textbf{Proof of Corollary \ref{Corq}:} Lemma 21.3 in van der Vaart
(1998) shows the pathwise differentiability of the quantile functional with an
influence function
\[
r_{\phi}(\alpha)=\frac{-\left\{  1(\alpha<\phi(\eta_{0}))-\tau\right\}  }%
{\dot{\eta}_{0}(\phi(\eta_{0}))}.
\]
That is, under the regularity conditions of the corollary, the quantile
functional $\eta_{0}\rightarrow\phi(\eta_{0})$ satisfies, for all $b\in
T(\eta_{0}),$%
\[
\lim_{t\rightarrow0}\frac{\phi(\eta_{t})-\phi(\eta_{0})}{t}=\mathbb{E}%
_{\eta_{0}}[r_{\phi}(\alpha)b(\alpha)].
\]
From Van der Vaart (1991) it follows that a necessary condition for the
quantile functional to be differentiable is%
\[
r_{\phi}(\alpha)-\phi(\eta_{0})=\int s(z)f_{z/\alpha}(z)d\mu(z).
\]
By Lemma \ref{Main} if the influence function of the functional is
discontinuous then the functional is not regularly identified. Since the
influence function of the quantile is not continuous, this proves the lemma.
$\blacksquare$\bigskip\

\noindent\textbf{Proof of Proposition \ref{PropABS}}: By substitution of
$f_{z/\alpha}(t_{1},t_{2})$ we obtain
\begin{align*}
\mathbb{E}\left[  \left.  s(Z)\right\vert \alpha\right]   &  =\int%
_{\mathcal{T}^{2}}s(t_{1},t_{2})f_{z/\alpha}(t_{1},t_{2})dt_{1}dt_{2}\\
&  =C\beta^{2}e^{2\alpha\beta}h(\alpha_{1}^{2},\alpha_{2}^{2}),
\end{align*}
where%
\[
h(u,v)=\int_{\mathcal{T}^{2}}s(t_{1},t_{2})\frac{1}{t_{1}^{3/2}t_{2}^{3/2}%
}s(u,v;t_{1})s(u,v;t_{2})dt_{1}dt_{2}%
\]
and
\[
s(u,v;t)=\exp\left(  -\frac{ut}{2}-\frac{v}{2t}\right)  ,\text{ }%
t\in\mathcal{T},\text{ }(u,v)\in(0,\infty).
\]
We check that the conditions for an application of the Leibniz's rule hold.
These conditions are

\begin{description}
\item[1.] The partial derivative $\partial^{m}s(u,v;t_{1})s(u,v;t_{2}%
)/\partial^{m}u$ exists and is a continuous function on an open neighborhood
$B$ of $(u,v),$ for a.s. $(t_{1},t_{2})\in\mathcal{T}^{2}.$

\item[2.] There is a positive function $h_{m}(t_{1},t_{2})$ such that%
\begin{equation}
\sup_{(u,v)\in B}\left\vert \frac{\partial^{m}s(u,v;t_{1})s(u,v;t_{2}%
)}{\partial^{m}u}\right\vert \leq h_{m}(t_{1},t_{2}) \label{dom}%
\end{equation}
and%
\begin{equation}
\int_{\mathcal{T}^{2}}s(t_{1},t_{2})\frac{1}{t_{1}^{3/2}t_{2}^{3/2}}%
h_{m}(t_{1},t_{2})dt_{1}dt_{2}<\infty. \label{int}%
\end{equation}

\end{description}

\noindent Simple differentiation and induction show that for any integer
$m\geq0$%
\[
\frac{\partial^{m}s(u,v;t_{1})s(u,v;t_{2})}{\partial^{m}u}=2^{-m}%
(-1)^{m}(t_{1}+t_{2})^{m}s(u,v;t_{1})s(u,v;t_{2}).
\]
Therefore, by monotonicity we can find $u^{\ast}$ and $v^{\ast}$ such that
(\ref{dom}) holds with
\[
h_{m}(t_{1},t_{2})=2^{-m}(t_{1}+t_{2})^{m}s(u^{\ast},v^{\ast};t_{1})s(u^{\ast
},v^{\ast};t_{2}).
\]
Furthermore, by $\mathbb{E}\left[  \left.  s(Z)\right\vert \alpha\right]
<\infty$ for all $\alpha$ in a local neighborhood (by local boundedness of
$r),$ and the boundedness of $\mathcal{T}$, condition (\ref{int}) holds. The
continuity of $h(u,v)$ is a special case of the previous arguments with $m=0$
(note the term $(t_{1}+t_{2})^{m}$ is one and the boundedness of $\mathcal{T}$
is not needed in this case). $\blacksquare$\bigskip

\noindent\textbf{Proof of Proposition \ref{PropBRC}}: Define
\begin{align*}
b(\alpha) &  =\mathbb{E}\left[  \left.  s(Y_{i}=1,X_{i})\right\vert \alpha
_{i}=\alpha\right]  \\
&  =\int1\left(  x^{\prime}\alpha\geq0\right)  s(1,x)dv_{X}(x).
\end{align*}
We prove that $b$ is continuous and by compactness of the sphere is therefore
uniformly continuous. Since the halfspaces $1\left(  x^{\prime}\alpha
\geq0\right)  $ and $1\left(  x^{\prime}\alpha_{0}\geq0\right)  $ intersect in
sets having surface measure of order $\left\vert \alpha-\alpha_{0}\right\vert
,$ it follows from the absolutely continuity of the angular component of $X$
that
\[
\left\vert b(\alpha)-b(\alpha_{0})\right\vert =O\left(  \left\vert
\alpha-\alpha_{0}\right\vert \right)  .
\]
When $x=(1,\tilde{x}),$ then%
\begin{align*}
b(\alpha)  & =\int1\left(  \tilde{x}^{\prime}\alpha_{2}\geq-\alpha_{1}\right)
s(1,1,\tilde{x})dv_{X}(\tilde{x}),\\
& =\int1\left(  u\geq-\alpha_{1}\right)  s_{\alpha_{2}}(u)f_{\alpha_{2}}(u)du,
\end{align*}
where $s_{\alpha_{2}}(u)=\mathbb{E}\left[  \left.  s(Y_{i}=1,1,\tilde{X}%
_{i})\right\vert \alpha_{2}^{\prime}\tilde{X}_{i}=u\right]  $ and
$f_{\alpha_{2}}$ denotes the density of $\alpha_{2}^{\prime}\tilde{X}_{i}.$
The absolute continuity in $\alpha_{1}$ follows from the integrability of
$s_{\alpha_{2}}(u)f_{\alpha_{2}}(u)$ and Royden (1968, Chapter 5).
$\blacksquare$\bigskip

\noindent\textbf{Proof of Corollary \ref{CorcdfBRC}}: The proof follows as in
Corollaries \ref{Corcdf} and \ref{Corq}. $\blacksquare$\bigskip

\noindent For a function $a\in L_{1}(\lambda)\cap L_{2}(\lambda),$ define the
Fourier transform $\hat{a}(t)=\int e^{it^{\prime}\alpha}a(\alpha)d\alpha,$
where $i=\sqrt{-1}.$ Use the notation
\[
\tilde{g}(p,x)=\int e^{ipy}g(y,x)dy,
\]
for the Fourier transform with respect to just the first argument (for
$g(\cdot,x)\in L_{1}(\lambda)\cap L_{2}(\lambda)).$ Define the norms%
\begin{equation}
\left\vert g\right\vert _{1,\rho}^{2}=\int_{\mathbb{S}^{d_{\alpha}-1}}%
\int_{\mathbb{R}}\left\vert \tilde{g}(p,x)\right\vert ^{2}(1+\left\vert
p\right\vert ^{2})^{\rho}dpdx\label{norm1}%
\end{equation}
and%
\begin{equation}
\left\vert g\right\vert _{\rho}^{2}=\int\left\vert \hat{g}(t)\right\vert
^{2}(1+\left\vert t\right\vert ^{2})^{\rho}dt.\label{Sobnorm}%
\end{equation}
The Sobolev space $H^{\rho}(\mathcal{A})$ is defined as the set of measurable
functions $g$ such that $\left\vert g\right\vert _{\rho}<\infty.$ \bigskip

\noindent\textbf{Proof of Proposition \ref{PropLRC}}: Define the score
operator $A:T(\eta_{0})\rightarrow L_{2}$
\[
Ab(z)=\frac{Rb\eta_{0}(z)}{f_{\eta_{0}}(z)}1(f_{\eta_{0}}(z)>0),
\]
where $R$ denotes the Radon transform
\[
Ra(y,x)=\int a(\alpha)1(y=x^{\prime}\alpha)d\alpha.
\]
Define $g(z)=s(z)f_{\eta_{0}}(z)$ and $a(\alpha)=b(\alpha)\eta_{0}(\alpha).$
Since $f_{\eta_{0}}(z)$ and $\eta_{0}$ are bounded, it follows that $g$ and
$a$ are in $L_{1}(\lambda)\cap L_{2}(\lambda).$ From the definition of
$Ra(y,x)$
\begin{equation}
\sup_{y,x}\left\vert Ra(y,x)\right\vert \leq\int\left\vert a(\alpha
)\right\vert d\alpha<\infty,\label{P2}%
\end{equation}
and since the supports of $\alpha$ and $X$ are bounded, the support of $Y$ is
also bounded and $Ra\in L_{2}(\lambda),$ so we can view $R:L_{2}%
(\lambda)\rightarrow L_{2}(\lambda).$

First, we show that if $s$ belongs to the closure of the range of $A,$ then
$g(z)=s(z)f_{\eta_{0}}(z)$ belongs to the closure of the range of $R.$ Indeed,
if $s_{n}$ is a sequence in the range of $A$ converging to $s$ in $L_{2},$
then $g_{n}=s_{n}f_{\eta_{0}}(z)\equiv Ra_{n}$ and clearly%
\[
\int\left\vert g_{n}(z)-g(z)\right\vert ^{2}dz\leq\int\left\vert
s_{n}(z)-s(z)\right\vert ^{2}f_{\eta_{0}}(z)dz\rightarrow0.
\]
Next, we shall show that any function $g$ in the closure of the range of $R$
will have an squared integrable weak derivative with respect to the first
argument $($in $y).$ By Theorem 2.4.1 in Ramm and Katsevich (1996) and
Assumption \ref{ALRC}(iii) it follows that $\left\vert g\right\vert _{1,\rho
}<\infty$ for $\rho=\rho_{0}+(d_{\alpha}-1)/2.$ While by well known results in
Fourier analysis, with $\partial_{y}g$ denoting the weak derivative with
respect to $y$%
\begin{align*}
\int_{\mathbb{S}^{d_{\alpha}-1}}\int\left\vert \widetilde{\partial_{y}%
g}(p,x)\right\vert ^{2}dpdx  &  \leq\int_{\mathbb{S}^{d_{\alpha}-1}}%
\int\left\vert p\right\vert ^{2}\left\vert \widetilde{g}(p,x)\right\vert
^{2}dpdx\\
&  \leq\int_{\mathbb{S}^{d_{\alpha}-1}}\int\left\vert \tilde{g}%
(p,x)\right\vert ^{2}(1+\left\vert p\right\vert ^{2})^{\rho}dpdx\\
&  <\infty,
\end{align*}
and similarly, by Cauchy-Schwarz
\begin{align*}
\int_{\mathbb{S}^{d_{\alpha}-1}}\int\left\vert \widetilde{\partial_{y}%
g}(p,x)\right\vert dpdx  &  \leq\int_{\mathbb{S}^{d_{\alpha}-1}}\int\left(
1+\left\vert p\right\vert ^{2}\right)  ^{1/2}\left\vert \widetilde{g}%
(p,x)\right\vert dpdx\\
&  \leq C\left(  \int_{\mathbb{S}^{d_{\alpha}-1}}\int(1+\left\vert
p\right\vert ^{2})^{1-\rho}dpdx\right)  ^{1/2}\\
&  <\infty,\text{ because }\rho>2.
\end{align*}
Thus $\widetilde{\partial_{y}g}(p,x)\in L_{1}(\lambda)\cap L_{2}(\lambda)$ and
by Plancherell's theorem $\partial_{y}g(\cdot)\in L_{2}(\lambda),$ as we claimed.

Define $\varphi(\cdot)=\partial_{y}g(\cdot)\in L_{2}(\lambda).$ We proceed to
verify the conditions of the dominated convergence theorem, see Lemma
\ref{Main}. First, we show that $g(y,x)$ is continuous in $y.$ Indeed, by the
bounded support assumption
\[
g(y,x)=\int_{-\infty}^{y}\varphi(u,x)dx
\]
is absolutely continuous in $y$ (see Royden 1968, Chapter 5).

Next, by independence of $\alpha_{i}$ and $X_{i},$%
\[
\mathbb{P}\left[  \left.  Y_{i}\leq y\right\vert X_{i}=x\right]
=\mathbb{P}\left[  x^{\prime}\alpha_{i}\leq y\right]  ,
\]
and taking derivatives we conclude $f_{\eta_{0}}(z)=\eta_{0,x}(y).$ Thus,
$f_{\eta_{0}}(z)$ is also continuous in $y$ by Assumption \ref{ALRC}(i).
Moreover,
\[
\inf_{\alpha\in N}\eta_{0,x}(x^{\prime}\alpha)\geq1/l(x)>0,
\]
which yields the continuity of $\alpha\rightarrow s(x^{\prime}\alpha,x)$ in
$N.$ Furthermore, by Cauchy-Schwarz and
\begin{align*}
\int\sup_{\alpha\in\Gamma_{0}}\left\vert s(x^{\prime}\alpha,x)\right\vert
f_{X}(x)dx  &  =\int\sup_{\alpha\in\Gamma_{0}}\left\vert g(x^{\prime}%
\alpha,x)\right\vert \sup_{\alpha\in\Gamma_{0}}\left\vert \frac{f_{X}%
(x)}{f_{\eta_{0}}(x^{\prime}\alpha,x)}\right\vert dx\\
&  \leq\left(  \int\left\vert \varphi(u,x)\right\vert ^{2}dudx\right)
^{1/2}\left(  \int\sup_{\alpha\in\Gamma_{0}}\left\vert \frac{f_{X}(x)}%
{f_{\eta_{0}}(x^{\prime}\alpha,x)}\right\vert ^{2}dx\right)  ^{1/2}\\
&  \leq C\int l^{2}(x)f_{X}(x)dx\\
&  \leq C.
\end{align*}
Thus, by dominated convergence $r$ must be continuous in $N$. $\blacksquare
$\bigskip

\noindent\textbf{Proof of Corollary \ref{CorcdfLRC}}: The proof follows as in
Corollaries \ref{Corcdf} and \ref{Corq}. $\blacksquare$\bigskip

\noindent\textbf{Proof of Proposition \ref{ATE}}: A necessary condition for a
reduced form functional $\phi(\eta_{0})=\mathbb{E}_{\eta_{0}}[r(\alpha)]$ to
be regularly identified is
\[
r(\alpha)-\phi(\eta_{0})=\int s(\alpha_{0}+\alpha_{1}x,x)dv_{X}(x),\text{
}\alpha=(\alpha_{0}^{\prime},\alpha_{1}^{\prime})=(\pi_{0},U_{2},\pi
_{1},\delta)^{\prime}.
\]
Thus, by Proposition \ref{PropLRC} $r(\alpha)$ must be continuous in $N.$
However, the influence function for the PPAME
\[
r_{PPAME}(\alpha)=1(\pi_{1}>0)1(\delta>0)+1(\pi_{1}<0)1(\delta<0)
\]
is discontinuous at the points $(p_{0},u_{2},0,d_{0})$ or $(p_{0},u_{2}%
,p_{1},0).$ Conclude that the PPAME is not regularly identified. As for AME,
by $\mathbb{E}\left[  \gamma^{2}\right]  <\infty$ this functional is
differentiable in the sense of van der Vaart (1991) with an influence function
$r_{AME}(\beta)=\pi_{1}/\delta.$ Since there is no continuous function that is
$\eta_{0}-$a.s equal to $r_{AME}(\beta)=\pi_{1}/\delta\ $when $(p_{0}%
,u_{2},p_{1},0)$ is a point in the interior of the support, we conclude that
the AME is not regularly identified. $\blacksquare$\bigskip

\noindent\textbf{Proof of Lemma \ref{ReguSem}}: By Lemma 25.34 in van der
Vaart (1998) the so-called score operator is given by
\[
Ab(z)=\mathbb{E}\left[  \left.  b(\alpha)\right\vert Z\right]  ,\text{ }b\in
T(\eta_{0})
\]
Thus, by the law of iterated expectations%
\begin{align*}
\mathbb{E}\left[  Ab(Z)s(Z)\right]   &  =\mathbb{E}\left[  b(\alpha
)s(Z)\right]  \\
&  =\mathbb{E}\left[  b(\alpha)\mathbb{E}\left[  \left.  s(Z)\right\vert
\alpha\right]  \right]  \\
&  =\mathbb{E}\left[  b(\alpha)\Pi_{\overline{T(\eta_{0})}}\mathbb{E}\left[
\left.  s(Z)\right\vert \alpha\right]  \right]  .
\end{align*}
In Lemma \ref{Main} we have shown that the functional $\eta_{0}\rightarrow
\phi(\eta_{0})=\mathbb{E}_{\eta_{0}}[r(\alpha)]$ is differentiable with
influence function%
\[
\chi(\alpha)=\Pi_{\overline{T(\eta_{0})}}r(\alpha).
\]
The lemma then follows from Theorem 3.1 in van der Vaart (1991).
$\blacksquare$\bigskip

\noindent\textbf{Proof of Lemma \ref{Bound}}: The sieve measure of
ill-posedness (cf. Blundell, Chen and Kristensen 2007) is%
\[
\tau_{D}=\sup_{b\in T(\eta_{0}),b\neq0}\frac{\left\Vert b\right\Vert
}{\left\Vert Ab\right\Vert }.
\]
Since $T(\eta_{0})=span(l_{\eta})$ and $\mathbb{E}_{\eta_{0}}\left[  l_{\eta
}(\alpha)l_{\eta}^{\prime}(\alpha)\right]  $ is the identity then
$b=\lambda^{\prime}l_{\eta}$ and $\left\Vert b\right\Vert ^{2}=\lambda
^{\prime}\lambda=\left\vert \lambda\right\vert ^{2},$ while $\left\Vert
Ab\right\Vert ^{2}=\lambda^{\prime}\mathbb{E}\left[  s_{0}(Z)s_{0}^{\prime
}(Z)\right]  \lambda.$ Thus,%
\begin{align*}
\tau_{D}^{2}  & =\sup_{\lambda\in\mathbb{R}^{D},\lambda\neq0}\frac{\left\vert
\lambda\right\vert ^{2}}{\lambda^{\prime}\mathbb{E}\left[  s_{0}%
(Z)s_{0}^{\prime}(Z)\right]  \lambda}\\
& =\frac{1}{\inf_{\lambda\in\mathbb{R}^{D},\left\vert \lambda\right\vert
=1}\lambda^{\prime}\mathbb{E}\left[  s_{0}(Z)s_{0}^{\prime}(Z)\right]
\lambda}\\
& =\frac{1}{\xi_{\min}(D)}.
\end{align*}
The bound then follows from Lemma 1 in Blundell, Chen and Kristensen (2007).
$\blacksquare$

\section{Appendix B: Further Results}

\subsection{Nonlinear RC}

\label{NRC}

In this section we describe a generic approach that can be used for generic
nonlinear RC models with continuous outcomes. We also illustrate how certain
invertible RC models are ruled out by our conditions. For the generic RC model
in (\ref{GRC}), the regularity condition reads
\begin{equation}
r(\alpha)-\phi(\eta_{0})=\mathbb{E}\left[  s(m(X_{i},\alpha),X_{i})\right]  .
\label{RGRC}%
\end{equation}
Again, the main difficulty in proving that the right hand side of (\ref{RGRC})
is continuous is that the score function $s(\cdot)$ is only known to be in
$L_{2}$ (thus, $s$ is potentially very discontinuous). To overcome this
difficulty, we resort to Fourier analysis and use the so-called Parseval's
identity (see Rudin 1987, pg. 187). To describe the method, assume $X$ is
absolutely continuous with density $f_{X}(x),$ and define%
\[
g(z)=s(z)f_{\eta_{0}}(z)\qquad\text{and}\qquad w(z,\alpha)=\frac{1\left(
y=m(x,\alpha)\right)  f_{X}(x)}{f_{\eta_{0}}(z)}1(f_{\eta_{0}}(z)>0).
\]
Note that $g\in L_{1}(\lambda),$ and since $f_{\eta_{0}}$ is bounded, also
$g\in L_{2}(\lambda).$ Let $\eta_{m,x}$ denote the density of $m(x,\alpha)$
when $\alpha$ has density $\eta_{0}.$ Under our conditions below,
$w(\cdot,\alpha)\in L_{1}(\lambda)\cap L_{2}(\lambda),$ and by Parseval's
identity, if $r$ satisfies (\ref{RGRC}) then
\begin{equation}
r(\alpha)-\phi(\eta_{0})=\int\hat{g}(t)\overline{\hat{w}(t,\alpha)}dt,
\label{Pars}%
\end{equation}
where, for a generic function $h\in L_{1}(\lambda),$ $\hat{h}(t)=(2\pi
)^{-d_{z}/2}\int e^{-it^{\prime}z}h(z)dz$ denotes the Fourier transform$,$
with $i=\sqrt{-1},$ $\overline{v}$ denotes the complex conjugate of $v$ and
\[
\overline{\hat{w}(t,\alpha)}=(2\pi)^{-d_{z}/2}\int\frac{f_{X}(x)}{\eta
_{m,x}(x^{\prime}\alpha)}e^{i(t_{1}m(x,\alpha)+t_{2}^{\prime}x)}dx.
\]
This integral representation is now amenable to our Lemma \ref{Main} under the
following assumption.

\begin{assumption}
\label{AGRC} (i) The vector $X$ is absolutely continuous with a bounded
density $f_{X}(\cdot)$; (ii) the density $\eta_{m,x}$ is continuous and
satisfies $\inf_{\alpha\in N}\eta_{m,x}(m(x,\alpha))>1/l(x)$ for an a.s.
positive measurable function $l(\cdot)$ such that $\mathbb{E}_{X}%
[l^{2}(X)]<\infty;$ (iii) the function $\alpha\rightarrow m(x,\alpha)$ is
continuous a.s. in $x;$ (iv) for all $\hat{g}$ satisfying (\ref{Pars})$,$
\begin{equation}
\int\left\vert \hat{g}(t)\right\vert \sup_{\alpha\in\Gamma_{0}}\left\vert
\overline{\hat{w}(t,\alpha)}\right\vert dt<\infty. \label{integ}%
\end{equation}

\end{assumption}

\begin{proposition}
\label{PropGRC}Under Assumption \ref{AGRC} and if $r\ $satisfies (\ref{RLRC}),
then $r(\cdot)$ must be continuous on $N.$\bigskip
\end{proposition}

\noindent\textbf{Proof of Proposition \ref{PropGRC}}: First, we need to check
that $g$ and $w(z,\alpha)$ are in $L_{1}(\lambda)\cap L_{2}(\lambda),$ so we
can apply Parseval's identity. From $s\in L_{2}$ and the definition of
$g(z)=s(z)f_{\eta_{0}}(z),$ it is clear that $g\in L_{1}(\lambda).$ Next, note%
\[
f_{\eta_{0}}(z)\leq\int_{\mathbb{R}^{d}}d\eta_{0}(\alpha)=1.
\]
Thus, $g$ also belongs to $L_{2}(\lambda).$ Furthermore, by independence of
$\alpha_{i}$ and $X_{i},$%
\[
\mathbb{P}\left[  \left.  Y_{i}\leq y\right\vert X_{i}=x\right]
=\mathbb{P}\left[  m(x,\alpha_{i})\leq y\right]  ,
\]
and taking derivatives we conclude $f_{\eta_{0}}(z)=\eta_{m,x}(y).$ Then, for
$p=1$ or $2,$
\begin{align*}
\int\left\vert w(z,\alpha)\right\vert ^{p}dz  &  =\int\left\vert \frac
{f_{X}(x)}{\eta_{m,x}(x^{\prime}\alpha)}\right\vert ^{p}dx\\
&  \leq\int l^{p}(x)\left\vert f_{X}(x)\right\vert ^{p}dx\\
&  \leq C\int l^{p}(x)f_{X}(x)dx\\
&  <\infty,
\end{align*}
because $f_{X}$ is bounded. Then, we can apply Parseval's identity and obtain
\[
r(\alpha)-\phi(\eta_{0})=\int\hat{g}(t)\overline{\hat{w}(t,\alpha)}dt.
\]
We now proceed to verify the conditions of Lemma \ref{Main} with $\hat
{g}(\cdot)$ playing the role of $s$ and $\overline{\hat{w}(t,\alpha)}$ that of
the conditional density. Note%
\[
\overline{\hat{w}(t,\alpha)}=(2\pi)^{-d_{z}/2}\int\frac{f_{X}(x)}{\eta
_{m,x}(m(x,\alpha))}e^{i(t_{1}m(x,\alpha)+t_{2}^{\prime}x)}dx.
\]
Under the conditions of the proposition the function $\alpha\rightarrow
\overline{\hat{w}(t,\alpha)}$ is continuous on $N$ since $\eta_{m,x}(\cdot)$
and $m(x,\cdot)$ are continuous and $\eta_{m,x}(m(x,\alpha))$ is bounded away
from zero on $N.$ Furthermore, the dominance condition holds from
(\ref{integ}). Conclude applying one more time dominated convergence under the
dominance condition Assumption \ref{AGRC}(iii). $\blacksquare$\bigskip\

\noindent Among the conditions of Assumption \ref{AGRC}, the most important
one is (\ref{integ}). We will see that in a class of invertible models this
condition fails to be satisfied. Consider the canonical monotonic nonseparable
model%
\[
Y_{i}=m(X_{i},\alpha_{i})
\]
with a scalar $\alpha_{i}$ and where $\alpha\rightarrow m(x,\alpha)$ is
strictly increasing with inverse $m^{-1}(y,x).$ Then, if we define
$s(Y_{i},X_{i})=1(m^{-1}(Y_{i},X_{i})\leq0),$ then the regularity condition of
Lemma \ref{Regu} is satisfied with $r(\alpha)=1(\alpha\leq0),$ proving that
the necessary condition for regular identification of the CDF at 0 (or at any
other point in fact) holds. In invertible models like this, regularity of CDFs
and quantiles is satisfied even in cases where $m$ is not known, but
identified. Our results do not apply to invertible models where heterogeneity
can be recovered as an identified function of observables.

To give a specific example, consider the model $Y_{i}=X_{i}+\alpha_{i},$ where
$s(Y_{i},X_{i})=1(Y_{i}\leq X_{i})$ solves (\ref{reg}) with $r(\alpha
)=1(\alpha\leq0),$ which is discontinuous at $0.$ This is of course an
unrealistic model, but the idea is simply to illustrate which of our
assumptions is key for the results to hold. In this example, Assumption
\ref{AGRC}(i-ii) is satisfied under mild conditions, since $\eta
_{m,x}(m(x,\alpha))=\eta_{0}(\alpha),$ but the integrability condition
(\ref{integ}) fails, since for $s(Y_{i},X_{i})=1(Y_{i}\leq X_{i})$
\begin{align*}
\int\left\vert \hat{g}(t)\right\vert \sup_{\alpha\in\Gamma_{0}}\left\vert
\overline{\hat{w}(t,\alpha)}\right\vert dt  &  =\inf_{\alpha\in\Gamma_{0}}%
\eta_{0}(\alpha)\int\left\vert \hat{g}(t_{1})\right\vert dt_{1}\\
&  =\infty,
\end{align*}
where $\hat{g}(t_{1})=\int1(\alpha\leq0)\eta_{0}(\alpha)e^{it_{1}\alpha
}d\alpha.$ Note that the discontinuity implies the lack of integrability.

\subsection{Identification under Kotlarski's Assumptions}

There is a growing literature in econometrics identifying the distribution of
latent variables by means of Kotlarski's Lemma (see Prakasa Rao (1983) for a
description of the method). In this setting we observe $Z=(Y_{1},Y_{2})$
satisfying
\begin{align*}
Y_{1}  &  =\alpha_{1}+\alpha_{2}\\
Y_{2}  &  =\alpha_{1}+\alpha_{3},
\end{align*}
where $\alpha=(\alpha_{1},\alpha_{2},\alpha_{3})^{\prime}$ is a vector of UH
with independent components, and (with some abuse of notation) Lebesgue
densities $\eta_{0j},$ for $j=1,2,3$. The density of the data is given by
\begin{align*}
f_{\eta_{0}}(y_{1},y_{2})  &  =\int1(y_{1}=\alpha_{1}+\alpha_{2}%
)1(y_{2}=\alpha_{1}+\alpha_{3})\eta_{01}(\alpha_{1})\eta_{02}(\alpha_{2}%
)\eta_{03}(\alpha_{3})d\alpha_{1}d\alpha_{2}d\alpha_{3}\\
&  =\int\eta_{02}(y_{1}-\alpha_{1})\eta_{03}(y_{2}-\alpha_{1})\eta_{01}%
(\alpha_{1})d\alpha_{1}.
\end{align*}
Consider a parametric submodel where $\eta_{02}$ and $\eta_{03}$ are known and
continuous. The model reduces then to our original setting where $f_{z/\alpha
}(z)=\eta_{02}(y_{1}-\alpha)\eta_{03}(y_{2}-\alpha)$ is known and continuous
in $\alpha$. If the dominance condition of Lemma \ref{Main} is satisfied, then
the CDF and quantiles of $\eta_{01}$ will be irregularly identified.

\subsection{Mixed Proportional Hazard Models}

The Mixed Proportional Hazard Model leads to a conditional density for
duration $Y$ given a vector of covariates $X$ given by
\[
f_{\eta_{0}}(y,x)=\int\phi(x)\psi(y)\alpha e^{-\phi(x)\Psi(y)\alpha}d\eta
_{0}(\alpha),
\]
where $\phi(x)$ is a transformation of covariates, $\Psi(y)$ is the baseline
cumulative hazard, with derivative $\psi,$ and $\alpha$ denotes UH. In
submodel where $\phi(x)$ and $\psi(y)$ are known, the model fits our original
formulation with $f_{z/\alpha}(z)=\phi(x)\psi(y)\alpha e^{-\phi(x)\Psi
(y)\alpha}$ known and continuous as a function of $\alpha$. Indeed, Horowitz
(1999) has established very slow rates of convergence (logarithmic) for the
CDF of $\alpha$, consistent with the irregular identification.

\subsection{Anatomy of the general problem}

The necessary condition for regular estimation in van der Vaart (1991) is
quite general, and in its abstract form reads as%
\[
\tilde{\psi}\in R(A^{\ast}),
\]
where $\tilde{\psi}$ is the so-called gradient, which for our original moment
functional is $\tilde{\psi}(\alpha)=r(\alpha)-\phi(\eta_{0}),$ and $A^{\ast}$
is the adjoint of the so-called score operator $A$. In many semiparametric
models, $A^{\ast}$ is a smoothing integral operator, in the sense that
\[
A^{\ast}s=\int s(z)k(z,\alpha)d\mu(z)
\]
is an operator from $L_{2}$ to $L_{2}(\eta_{0})$ with a kernel function $k$
such that $\alpha\rightarrow k(z,\alpha)$ is smooth, at least for some
submodel. We expect our results to be potentially applicable in this general setting.

\subsection{Further Simulation Results}

We report here further results for estimation of quantiles in the Mixed Logit
Model. The setting is that of the Monte Carlo section, the only different
being that other quantile levels $\tau$ different from the median ($\tau=0.5$)
are considered. Table 3 report the RMSE. We observe that, as expected, the
RMSE at more extreme quantiles are larger than those for the median. Again,
the gains from regularization are substantial, particularly for large values
of $R.$\newpage

\begin{center}
\textbf{Table 3}. RMSE for $\tau-$Quantiles of marginals of UH\newline%
\begin{tabular}
[c]{ccccccc}\hline\hline
$\tau$ & $n$ & $D$ & RMSEQ1 & RMSEQ1-PCR & RMSEQ2 & RMSEQ2-PCR\\\hline
0.25 & 100 & 25 & 1.6739 & 0.9308 & 1.7238 & 0.7270\\
0.25 & 500 & 25 & 1.3247 & 0.8674 & 1.3270 & 0.6305\\
0.25 & 1000 & 25 & 1.0596 & 0.8075 & 1.1569 & 0.6250\\
0.25 & 100 & 100 & 1.8369 & 0.6098 & 1.8217 & 0.6608\\
0.25 & 500 & 100 & 1.4038 & 0.4929 & 1.3763 & 0.5504\\
0.25 & 1000 & 100 & 1.3153 & 0.4832 & 1.2041 & 0.5036\\
0.25 & 100 & 500 & 1.8075 & 0.5893 & 1.8696 & 0.6275\\
0.25 & 500 & 500 & 1.4529 & 0.4520 & 1.4698 & 0.4894\\
0.25 & 1000 & 500 & 1.2954 & 0.4444 & 1.2481 & 0.4500\\\hline
0.75 & 100 & 25 & 1.5719 & 0.9803 & 1.6573 & 0.9928\\
0.75 & 500 & 25 & 1.1938 & 0.8077 & 1.3158 & 0.7953\\
0.75 & 1000 & 25 & 0.8941 & 0.7045 & 1.1489 & 0.7525\\
0.75 & 100 & 100 & 1.8192 & 0.8616 & 1.7989 & 0.8099\\
0.75 & 500 & 100 & 1.2178 & 0.6029 & 1.1809 & 0.5936\\
0.75 & 1000 & 100 & 0.8947 & 0.5495 & 0.9476 & 0.5358\\
0.75 & 100 & 500 & 1.9017 & 0.8107 & 1.9402 & 0.8329\\
0.75 & 500 & 500 & 1.2381 & 0.5666 & 1.2324 & 0.5467\\
0.75 & 1000 & 500 & 0.9606 & 0.4885 & 0.9533 & 0.5102\\\hline
\end{tabular}

\end{center}

\newpage%

%

\end{document}